\documentclass[amsmath, pra, twocolumn, showpacs, floatfix]{revtex4-1}
\usepackage{graphicx}
\usepackage{dsfont}

\begin{document}   

\title{Nanofiber-based all-optical switches}
 
\author{Fam Le Kien}

\affiliation{Wolfgang Pauli Institute, Oskar Morgensternplatz 1, 1090 Vienna, Austria}

\author{A. Rauschenbeutel} 

\affiliation{Vienna Center for Quantum Science and Technology, Institute of Atomic and Subatomic Physics, Vienna University of Technology, Stadionallee 2, 1020 Vienna, Austria}

\date{\today}

\begin{abstract}
We study all-optical switches operating on a single four-level atom with the $N$-type transition configuration 
in a two-mode nanofiber cavity with a significant length (on the order of $20$ mm) and a moderate finesse (on the order of 300) under the electromagnetically induced transparency (EIT) conditions. 
In our model, the gate and probe fields are the quantum nanofiber-cavity fields excited by weak classical light pulses, and the parameters of the $D_2$ line of atomic cesium are used. 
We examine two different switching schemes. The first scheme is based on the effect of the presence of a photon in the gate mode on the EIT of the probe mode.
The second scheme is based on the use of EIT to store a photon of the gate mode in the population of an appropriate atomic level, which leads to
the reduction of the transmission of the field in the probe mode. We investigate the dependencies of the switching contrast on various parameters, such as the cavity length, the mirror reflectivity, and the detunings and powers of the cavity driving field pulses. For a nanofiber cavity with fiber radius of 250 nm,  cavity length of 20 mm, and  cavity finesse of 313 and a cesium atom at a distance of 200 nm from the fiber surface, we numerically obtain a switching contrast on the order of about 67\% for the first scheme and of about 95\% for the second
scheme. These switching operations require small mean numbers of photons in the nanofiber cavity gate and probe modes. 
\end{abstract}

\maketitle

\section{Introduction}
\label{sec:introduction}

Cavity quantum electrodynamics (QED) plays a central role in studies of fundamental quantum processes of interaction between light and atoms \cite{CQED}. 
When a single mode of the light field is selected by a high-finesse optical microcavity, the coherent interaction between the field and an atom in the cavity is significantly enhanced by the confinement of the field in the small mode volume of the cavity. Due to the discrete nature of the cavity mode structure, the transfer of quantum states between the atom and the cavity field is intrinsically reversible. Therefore, an atom coupled to an optical cavity field can serve as a node for a quantum network \cite{Rempe1999, Rempe2000, Rempe2002, Kimble2004, Kuhn, Rempe2007, Rempe2008, Rempe2009}, where quantum information is generated, stored, processed, and transmitted through. Controlled generation of single photons from such a system has been realized \cite{Rempe2002, Kimble2004, Kuhn}. 
Atom-photon entanglement \cite{Rempe2007} and photon-photon entanglement \cite{Rempe2009} in an optical cavity have been demonstrated. 
Cavity QED enables the realization of remarkable nonlinear optical
phenomena such as photon blockade, photon-induced tunneling, and all-optical switching, in which individual photons strongly interact with each other. 

In photon blockade, the transmission of one photon through a system hinders the transmission of subsequent photons \cite{Imamoglu97}. Meanwhile, in photon-induced tunneling,
the opposite behavior is observed, that is, the probability of admitting subsequent photons increases \cite{Faraon11}. Photon blockade and photon-induced tunneling are caused by
the anharmonicity of the energy levels of the coupled atom-cavity system. 
Photon blockade and photon-induced tunneling in a cavity with a two-level emitter have been reported \cite{Birnbaum05,Dayan08,Liew10,Majumdar12,Kubanek08,Faraon08,Faraon11}. Photon blockade based on the coupling of four-level quantum emitters to a cavity has been studied \cite{Imamoglu97,Werner99,Rebic99,Gheri99,Greentree00,Rebic02a,Rebic02b,Bajcsy13}.

In an all-optical switch, one light beam can fully control another light beam.
It has been shown that the transmission of light through an ensemble of atoms under conditions of electromagnetically induced transparency (EIT) in an optical cavity \cite{Lukin1998,review,Muecke2010,Suzuki11,Albert11,Dantan12} may be controlled with few photons and even by the electromagnetic vacuum field \cite{Suzuki11}.
All-optical switching in a four-level system has been studied \cite{Schmidt96,Harris98,Bajcsy09,Albert11,Dantan12,Chen13}.
All-optical switching at the level of a few hundred photons in an ensemble of four-level atoms within a hollow fiber has been demonstrated \cite{Bajcsy09}. 
The cavity QED version  \cite{Imamoglu97} of an all-optical few-photon switch \cite{Chang07} based on EIT in an ensemble of four-level atoms has been achieved \cite{Chen13}. 
Single-photon switches operating on a quantum dot in a cavity \cite{Volz12}, on a single atom coupled to a fiber-coupled, chip-based microresonator \cite{Shomroni14}, 
or on Rydberg blockade in an ensemble of atoms \cite{Baur14} have been realized.
Switching of light by a single emitter in a cavity \cite{Thompson92}, near a waveguide \cite{Chang07,Shen07}, or in a tightly focused laser beam \cite{Hwang09}  has also been studied.

Similar to microcavities, vacuum-clad silica-core fibers with diameters smaller than the wavelength of light can tightly confine the field. 
Such thin fibers are called nanofibers. A nanofiber can be produced as the waist of a tapered optical fiber \cite{Mazur's Nature, Birks}. The adiabatic tapering technique \cite{taper} allows one to 
match the mode of a conventional single-mode optical fiber with the mode of the subwavelength-diameter tapered waist region, thus ensuring high transmission and integrability of the device. 
In a nanofiber, the original core is, due to tapering, almost vanishing. The refractive indices that determine the guiding properties of the nanofiber are the refractive index of the original silica clad and the refractive index of the surrounding vacuum. The nanofiber field is an evanescent wave in the cross-section plane and propagates along the fiber. 

A nanofiber cavity can be obtained by combining the nanofiber technique with the
fiber-Bragg-grating (FBG) cavity technique \cite{fibercavity,cavityspon,cavitytrap,ramancavity,rubidium,milling,Ding11,Fam12,Wuttke12,Nayak13,Sadgrove13,Nayak14,Yalla14,Aoki15}. In such a system, the atom-field interaction is enhanced by the confinement of the field in the fiber cross-section plane and in the space between the built-in FBG mirrors. The output field is in guided modes and can therefore be transmitted over long distances for communication purposes. Various applications of nanofiber cavities have been studied \cite{fibercavity,milling,cavityspon,cavitytrap,ramancavity,rubidium,Yalla14,Aoki15}. It has been shown that a nanofiber cavity with a large length (on the order of 1--10 cm) and a moderate finesse (less than 1000) can  substantially enhance the channeling of emission from an atom into the nanostructure \cite{cavityspon}. The deterministic generation of a single guided photon has been studied \cite{ramancavity}.
The controlled generation of entangled guided photons from an atom in a nanofiber cavity has been investigated \cite{rubidium}. 
A significant enhancement of the spontaneous emission rate into the nanofiber guided modes has been demonstrated for single quantum dots in a nanofiber cavity \cite{Yalla14}.
Very recently, a nanofiber cavity with a single trapped atom in the strong-coupling regime has been demonstrated and
the vacuum Rabi splitting has been observed \cite{Aoki15}. The results of Ref.~\cite{Aoki15} make the study of nanofiber cavity quantum electrodynamics very timely and attractive.

In this paper, we study all-optical switches operating on a single four-level atom with the $N$-type transition configuration \cite{Imamoglu97,Schmidt96,Harris98,Bajcsy09,Albert11,Dantan12,Chen13}
in a two-mode nanofiber cavity with a significant length and a moderate finesse under the EIT conditions. We consider the case where both the gate field and the target field are the quantum cavity fields excited by weak classical pulses. We examine two different schemes where the switching occurs due to different mechanisms. 

The paper is organized as follows. In Sec.~\ref{sec:model}, we describe the model and present the basic equations. 
In Sec.~\ref{sec:dressed}, we present the analytical expressions for the dressed states of the coupled atom-cavity system. 
In Sec.~\ref{sec:option1}, we study the possibility to switch a cavity mode under the EIT conditions using the effect of the presence of a photon in the other mode on the EIT. 
In Sec.~\ref{sec:option2}, we investigate the possibility to switch a cavity mode by storing a photon of the other mode in the population of an appropriate atomic level. Our conclusions are given in Sec.~\ref{sec:summary}.

\section{Model}
\label{sec:model}

\begin{figure}[tbh]
\begin{center}
\includegraphics{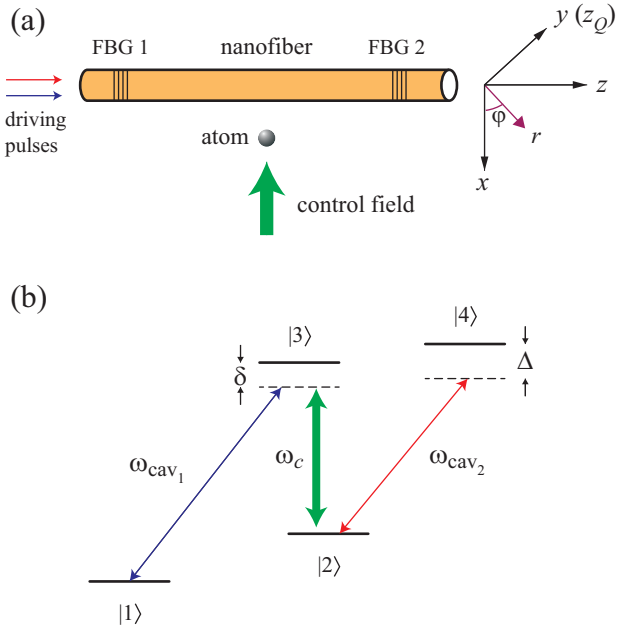}
\end{center}
\caption{(Color online) (a) Atom near a nanofiber with two FBG mirrors. The atom lies in the $zx$ plane, with $y$ being the quantization axis and $z$ being the fiber axis. (b) Scheme of energy levels and transitions of the atom. In our numerical calculations, we use the sublevels $|1\rangle=|6S_{1/2},F=3, M= 3\rangle$, $|2\rangle=|6S_{1/2},F=4, M= 4\rangle$, $|3\rangle=|6P_{3/2},F'=4, M'= 4\rangle$, and $|4\rangle=|6P_{3/2},F'=5, M'= 5\rangle$ of the $D_2$ line of atomic cesium. Two cavity quantum fields with resonant frequencies $\omega_{\mathrm{cav}_1}$ and $\omega_{\mathrm{cav}_2}$ couple the atomic transitions $|3\rangle\leftrightarrow|1\rangle$ and $|4\rangle\leftrightarrow|2\rangle$, respectively, with the detunings $\delta$ and $\Delta$, respectively. The cavity is driven by two weak classical guided fields. A strong external classical field of frequency $\omega_c$ is applied to the atomic transition $|3\rangle\leftrightarrow|2\rangle$. The detunings of the classical cavity-driving and atom-control fields are not shown. The external control field is linearly polarized along the $y$ axis, while the cavity guided modes are quasilinearly polarized along the $x$ axis.
}
\label{fig1}
\end{figure}

We consider a single four-level atom with the $N$-type transition configuration in a two-mode nanofiber cavity (see Fig.~\ref{fig1}).
The energy levels of the atom are labeled by the index $j=1,2,3,4$.
The corresponding basis internal states of the atom are denoted as $|j\rangle$, with the associated energies $\hbar\omega_j$.
The angular frequency of the atomic transition $|j\rangle\leftrightarrow|j'\rangle$ is $\omega_{jj'}=\omega_{j}-\omega_{j'}$.
The cavity is formed by a nanofiber with two built-in FBG mirrors \cite{milling,Ding11,Fam12,Wuttke12,Nayak13,Sadgrove13,Nayak14,Yalla14,Aoki15}.
We consider two cavity modes whose resonant frequencies $\omega_{\mathrm{cav}_1}$ and $\omega_{\mathrm{cav}_2}$ are near resonance with the atomic
transition frequencies $\omega_{31}$ and $\omega_{42}$, respectively. We label these cavity modes by the index $\nu=1,2$.
The cavity is driven by two weak classical guided fields of frequencies $\omega_{p_1}$ and $\omega_{p_2}$, 
which excite cavity modes 1 and 2, respectively. 
The cavity quantum fields in modes 1 and 2 couple the atomic transitions $|3\rangle\leftrightarrow|1\rangle$ and $|4\rangle\leftrightarrow|2\rangle$, 
respectively, with the coupling coefficients $g_1$ and $g_2$, respectively.
A strong external classical field with the complex amplitude $\boldsymbol{\mathcal{E}}_c$ and the angular frequency $\omega_c$ is applied to the atomic transition $|3\rangle\leftrightarrow|2\rangle$. 
The corresponding Rabi frequency is $\Omega_c=\mathbf{d}_{32}\cdot\boldsymbol{\mathcal{E}}_c/\hbar$, where $\mathbf{d}_{32}$ is the dipole matrix element
for the atomic transition $|3\rangle\leftrightarrow|2\rangle$.
The transitions $|4\rangle\leftrightarrow|3\rangle$, $|4\rangle\leftrightarrow|1\rangle$, and $|2\rangle\leftrightarrow|1\rangle$  are not allowed within the electric-dipole approximation.

We use the Cartesian coordinate system $\{x,y,z\}$ and the associated cylindrical coordinate system $\{r,\varphi,z\}$, with $z$ being the fiber axis. 
We assume that the atom lies in the $zx$ plane, that is, the position of the atom is $\{ x,0,z\}$. 
We call the axes $x$ and $y$ the major and minor principal axes, respectively.
To be specific, we use the transitions between the Zeeman sublevels of the $D_2$ line of atomic cesium in our numerical calculations. 
In order to specify the atomic states, we use the minor principal axis $y$ as the quantization axis $z_Q$.
We use the atomic states $|1\rangle=|6S_{1/2},F=3, M= 3\rangle$, $|2\rangle=|6S_{1/2},F=4, M= 4\rangle$,
$|3\rangle=|6P_{3/2},F'=4, M'= 4\rangle$, and $|4\rangle=|6P_{3/2},F'=5, M'= 5\rangle$. Here, $F$ (and $F'$) and $M$ (and $M'$) denote the hyperfine and magnetic sublevels, respectively. The effects of other Zeeman sublevels are removed by applying an external magnetic field.

\subsection{Effective Hamiltonian}
\label{subsec:Hamiltonian}

In the electric-dipole and rotating-wave approximations, the Hamiltonian of the atom-cavity system is
\begin{eqnarray}\label{1}
H&=&\hbar\sum_{j=1}^4\omega_j\sigma_{jj}+\hbar\sum_{\nu=1,2}\omega_{\mathrm{cav}_\nu} a_\nu^\dagger a_\nu\nonumber\\
&&\mbox{}-i\hbar (g_1a_1\sigma_{31}-g_1^*a_1^\dagger\sigma_{13})
-i\hbar (g_2a_2\sigma_{42}-g_2^*a_2^\dagger\sigma_{24})\nonumber\\
&&\mbox{}-\frac{\hbar}{2}(\Omega_c e^{-i\omega_ct}\sigma_{32}+\Omega_c^*e^{i\omega_ct}\sigma_{23})\nonumber\\
&&\mbox{}+\hbar\sum_{\nu=1,2}(\mathcal{E}_{p_\nu}e^{-i\omega_{p_\nu}t}a_\nu^\dagger+\mathcal{E}_{p_\nu}^*e^{i\omega_{p_\nu}t}a_\nu).
\end{eqnarray}
Here, $\sigma_{jj'}=|j\rangle\langle j'|$ are the atomic operators, 
$a_\nu$ and $a_\nu^\dagger$ are the photon annihilation and creation operators, respectively, 
and $\mathcal{E}_{p_\nu}$ characterize the strengths of the classical pump fields for the cavity. 

We use a rotating coordinate frame described by the unitary operator $U=e^{-iWt}$, where
\begin{eqnarray}\label{2}
W&=&\omega_{p_1}a_1^\dagger a_1+\omega_{p_2}a_2^\dagger a_2+\omega_1\sigma_{11}+(\omega_{p_1}-\omega_c+\omega_1)\sigma_{22}\nonumber\\
&&\mbox{}+(\omega_{p_1}+\omega_1)\sigma_{33}+(\omega_{p_1}+\omega_{p_2}-\omega_c+\omega_1)\sigma_{44}.
\end{eqnarray} 
Then, the original Hamiltonian \eqref{1} changes into the effective Hamiltonian
\begin{eqnarray}\label{3a}
\mathcal{H}&=&U^\dagger H U+i\hbar\frac{dU^\dagger}{dt}U.
\end{eqnarray}
We find $\mathcal{H}=\mathcal{H}_0+\mathcal{H}_P$, where
\begin{eqnarray}\label{5}
\mathcal{H}_0&=&\hbar\sum_{j=2}^4(\delta_j-\Delta_{\mathrm{cav}_1})\sigma_{jj}-\hbar\Delta_{\mathrm{cav}_2}\sigma_{44}\nonumber\\
&&\mbox{}-\hbar\sum_{\nu=1,2}\Delta_{\mathrm{cav}_\nu} a_\nu^\dagger a_\nu-i\hbar (g_1a_1\sigma_{31}-g_1^*a_1^\dagger\sigma_{13})\nonumber\\
&&\mbox{}     -i\hbar (g_2a_2\sigma_{42}-g_2^*a_2^\dagger\sigma_{24})
-\frac{\hbar}{2}(\Omega_c\sigma_{32}+\Omega_c^*\sigma_{23}),\nonumber\\
\mathcal{H}_P&=&\hbar\sum_{\nu=1,2}(\mathcal{E}_{p_\nu}a_\nu^\dagger+\mathcal{E}_{p_\nu}^*a_\nu).
\end{eqnarray}
Here, we have introduced the notations
\begin{equation}\label{4}
\begin{split}
\Delta_{\mathrm{cav}_\nu}&=\omega_{p_\nu}-\omega_{\mathrm{cav}_\nu},\\ 
\delta_2&=\omega_{21}-\omega_{\mathrm{cav}_1}+\omega_c,\\
\delta_3&=\omega_{31}-\omega_{\mathrm{cav}_1},\\
\delta_4&=\omega_{41}-\omega_{\mathrm{cav}_1}-\omega_{\mathrm{cav}_2}+\omega_c.
\end{split}
\end{equation}
It is also convenient to use the notations $\delta=\omega_{31}-\omega_{\mathrm{cav}_1}$ and $\Delta=\omega_{42}-\omega_{\mathrm{cav}_2}$.
In terms of these notations, we have $\delta_3=\delta$ and $\delta_4=\Delta+\delta_2$. 

The effects of the cavity damping and atomic decay can be taken into account by adding to \eqref{5} a non-Hermitian term
\begin{equation}\label{6}
\mathcal{H}_{\mathrm{damp}}=-\frac{i\hbar}{2}\sum_{\nu=1,2}\kappa_\nu a_\nu^\dagger a_\nu-\frac{i\hbar}{2}(\gamma_3\sigma_{33}+\gamma_4\sigma_{44}+\gamma_{2\mathrm{deph}}\sigma_{22}).
\end{equation}
Here, the coefficients $\kappa_\nu$ are the cavity damping rates for the cavity modes $\nu$, the coefficients $\gamma_3$ and $\gamma_4$ are the linewidths of the levels $|3\rangle$ and $|4\rangle$, respectively,
and the coefficient $\gamma_{2\mathrm{deph}}$ is twice the dephasing rate for the lower-level coherence.

The master equation for the density operator $\rho$ of the atom-cavity system is 
\begin{equation}\label{45}
\dot{\rho}=-\frac{i}{\hbar}[\mathcal{H},\rho]-\frac{i}{\hbar}(\mathcal{H}_{\mathrm{damp}}\rho-\rho\mathcal{H}^\dagger_{\mathrm{damp}})+\mathcal{J}\rho,
\end{equation}
where the operator
\begin{equation}\label{47}
\begin{split}
\mathcal{J}\rho&=\sum_{\nu=1,2}\kappa_\nu a_\nu\rho a_\nu^\dagger+\gamma_{31}\sigma_{13}\rho\sigma^\dagger_{13}
+\gamma_{32}\sigma_{23}\rho\sigma^\dagger_{23}\\
&\quad +\gamma_{42}\sigma_{24}\rho\sigma^\dagger_{24}+\gamma_{2\mathrm{deph}}\sigma_{22}\rho\sigma_{22}
\end{split}
\end{equation}
describes the jump. 
Here, the parameters $\gamma_{31}$, $\gamma_{32}$, and $\gamma_{42}$ are the spontaneous emission rates for the transitions $|3\rangle\leftrightarrow|1\rangle$,
$|3\rangle\leftrightarrow|2\rangle$, and $|4\rangle\leftrightarrow|2\rangle$, respectively.
In general, we have $\gamma_3\geq \gamma_{31}+\gamma_{32}$ and $\gamma_4\geq \gamma_{42}$.
In our numerical calculations, we have $\gamma_3>\gamma_{31}+\gamma_{32}$ and $\gamma_4=\gamma_{42}$. 
The inequality $\gamma_3>\gamma_{31}+\gamma_{32}$ is a consequence of the fact that, in the case of atomic cesium, the population of the upper level $|3\rangle=|6P_{3/2},F'=4, M'= 4\rangle$ can decay not only to the lower levels $|1\rangle=|6S_{1/2},F=3, M= 3\rangle$ and $|2\rangle=|6S_{1/2},F=4, M= 4\rangle$ but 
also to the other lower levels which are outside the working level configuration and, therefore, are not shown in Fig.~\ref{fig1}. The equality $\gamma_4=\gamma_{42}$ is a consequence of the fact that,
for atomic cesium, the population of the upper level $|4\rangle=|6P_{3/2},F'=5, M'= 5\rangle$ can decay only to the lower level $|2\rangle=|6S_{1/2},F=4, M= 4\rangle$.

\subsection{Nanofiber cavity and coupling coefficients}
\label{subsec:cavity}

We describe the nanofiber cavity field and derive the expressions for the atom-cavity coupling coefficients.
The nanofiber cavity is a nanofiber with two built-in FBG mirrors (see Fig. \ref{fig1}). The guided field in the nanofiber cavity is reflected back and forth between the FBG mirrors. The nanofiber has a cylindrical silica core of radius $a$ and of refractive index $n_1=1.45$ and an infinite vacuum clad of refractive index $n_2=1$. In view of the very low losses of silica in the wavelength range of interest, we neglect material absorption. We also neglect the effects of the surface-induced potential, the surface roughness, and the phonon heating on the atom.

In order to describe the field quantum mechanically, we follow the continuous-mode field quantization procedures presented in \cite{Loudon}. First, we temporally neglect the presence of the FBG mirrors. We assume that the single-mode condition \cite{fiber books} is satisfied for a finite bandwidth around the characteristic atomic transition frequency $\omega_0$. 
We label each fundamental guided mode HE$_{11}$ with a frequency $\omega$ in this bandwidth by an index $\mu=(\omega,f,l)$, 
where $f=+,-$ denotes the forward or backward propagation direction
and $l=+,-$ denotes the counterclockwise or clockwise rotation of the transverse component of the polarization with respect to the fiber axis $z$. In the interaction picture, the quantum expression for the electric positive-frequency component 
$\mathbf{E}^{(+)}_{\mathrm{gyd}}$ of the nanofiber guided field is \cite{cesium decay}
\begin{equation}\label{7}
\mathbf{E}^{(+)}_{\mathrm{gyd}}=i\sum_{\mu}\sqrt{\frac{\hbar\omega\beta'}{4\pi\epsilon_0}}
\;a_{\mu}\mathbf{e}^{(\mu)}e^{-i(\omega t-f\beta z-l\varphi)}.
\end{equation}
Here, $\mathbf{e}^{(\mu)}=\mathbf{e}^{(\mu)}(r,\varphi)$ is the profile function of the guided mode $\mu$ in the classical problem, $a_{\mu}$ is the corresponding photon annihilation operator, 
$\sum_{\mu}=\sum_{fl}\int_0^{\infty}d\omega$ is the summation over the guided modes,
$\beta$ is the longitudinal propagation constant, and $\beta'$ is the derivative of $\beta$
with respect to $\omega$. The propagation constant $\beta$ is determined by the
fiber eigenvalue equation \cite{fiber books}. The operators $a_{\mu}$ and $a_{\mu}^\dagger$ satisfy the continuous-mode bosonic commutation rules $[a_{\mu},a_{\mu'}^\dagger]=\delta(\omega-\omega')\delta_{ff'}\delta_{ll'}$. The explicit expression for the mode function $\mathbf{e}^{(\mu)}$ is given
in Refs. \cite{cesium decay,fiber books}. 
The normalization of the mode function is given by the condition 
\begin{equation}\label{g5}
\int _{0}^{2\pi}d\varphi\int _{0}^{\infty}n_{\mathrm{ref}}^2\,|\mathbf{e}^{(\mu)}|^2r\,dr=1.
\end{equation}
Here, $n_{\mathrm{ref}}(r)=n_1=1.45$ for $r<a$, and $n_{\mathrm{ref}}(r)=n_2=1$ for $r>a$.
 
Next, we take into account the effect of the FBG mirrors on the mode functions. 
We assume that the two FBG mirrors are identical, having the same complex reflection and transmission coefficients $R$ and $T$, respectively, for the guided modes in a broad bandwidth around the characteristic atomic transition frequency $\omega_0$. In general, we have $|R|^2+|T|^2\leq 1$, where the equality (inequality) occurs for lossless (lossy) gratings. Without loss of essential physics, we assume that the gratings are lossless, that is, $|R|^2+|T|^2=1$. Let the mirrors be separated by a distance $L$, from 
the point $z=-L/2$ to the point $z=L/2$. The mode functions of the guided modes are modified by the
presence of the mirrors. We assume that the FBG mirrors do not change the polarization of the field in the reflection and transmission. 
The forms of the cavity-modified mode functions are obtained, as usual in the Fabry-P\'{e}rot theory, by summing up the geometric series of the contributions 
of multiple reflections of the field from the mirrors
\cite{Martini,Bjork,Cook}. Inside the cavity, the mode functions of the cavity-modified guided modes are given by
\begin{eqnarray}\label{8}
\tilde{\mathbf{e}}^{(\omega,+,l)}&=&\mathbf{e}^{(\omega,+,l)}\frac{T}{1-R^2e^{2i\beta L}}
+\mathbf{e}^{(\omega,-,l)}\frac{TR e^{i\beta(L-2z)}}{1-R^2e^{2i\beta L}},
\nonumber\\
\tilde{\mathbf{e}}^{(\omega,-,l)}&=&\mathbf{e}^{(\omega,-,l)}\frac{T}{1-R^2e^{2i\beta L}}
+\mathbf{e}^{(\omega,+,l)}\frac{TR e^{i\beta (L+2z)}}{1-R^2e^{2i\beta L}},
\nonumber\\
\end{eqnarray}
and, hence, the electric positive-frequency component of the field in the cavity-modified guided modes is
\begin{equation}\label{9}
\mathbf{E}^{(+)}_{\mathrm{cav}}=i\sum_{\mu}\sqrt{\frac{\hbar\omega\beta'}{4\pi\epsilon_0}}
\;a_{\mu}\tilde{\mathbf{e}}^{(\mu)}e^{-i(\omega t-f\beta z-l\varphi)}.
\end{equation}

We emphasize that the quantization scheme presented above has two steps, expressed by Eqs.~\eqref{7} and \eqref{9}.
Equation \eqref{7} describes the quantization of the guided field in the absence of the FBG mirrors.
Equation \eqref{9}, which contains the cavity-modified mode functions \eqref{8}, describes the quantization of the guided field in the FBG cavity.
The effect of the FBG mirrors on the mode structure and the state density is ignored in Eq.~\eqref{7} but is partially accounted for in Eq.~\eqref{9}. 
The two-step quantization scheme described above is an approximation. Indeed, the FBG mirrors may lead to the coupling between different guided modes
and between guided modes and radiation modes of the bare fiber. The coupling between guided modes leads to mode mixing that may result
in polarization changing and birefringence.
The coupling between guided modes and radiation modes leads to losses. When the mode mixing and radiative losses, produced by the FBG mirrors, are not significant, the two-step quantization scheme holds.
Otherwise, a numerical method based on the Maxwell equations, the coupled-mode theory, or the transfer matrix for a fiber with FBG mirrors must be used. We assume in this paper that the mode coupling produced by the FBG mirrors is not serious and therefore the two-step quantization scheme can be used.
We note that the mode matching between the cavity field and the guided modes of the nanofiber is very good in the experiments \cite{milling,Ding11,Wuttke12,Nayak13,Sadgrove13,Nayak14,Yalla14,Aoki15}.

The resonant frequencies of the nanofiber cavity are determined by the minima of the absolute value of the denominator $1-R^2e^{2i\beta L}$ in Eqs.~\eqref{8}, that is, by the condition
$\beta L+\phi_R=n\pi$, where $\phi_R$ is the phase of the reflection coefficient $R$ and $n$ is an integer number characterizing the order of the resonance.
We assume that $|R|^2\simeq 1$. In the single-mode regime, the electric positive-frequency component of the field in a single excited nanofiber cavity mode $\alpha$ is given, for quasicircular polarization, by the expression
\begin{eqnarray}\label{10}
\mathbf{E}^{(+)}_{\alpha}&=&i\sqrt{\frac{\hbar\omega_{\mathrm{cav}}}{\epsilon_0 L}}
\;a
\big[\big(\hat{\mathbf{r}}e_r+l\hat{\boldsymbol{\varphi}}e_\varphi\big)\cos\beta_{\mathrm{cav}}(z-z_0)
\nonumber\\&&\mbox{}
+i\hat{\mathbf{z}}e_z\sin\beta_{\mathrm{cav}}(z-z_0)
\big]e^{il\varphi}e^{-i\omega_{\mathrm{cav}} t}
\end{eqnarray}
and, for quasilinear polarization, by the expression 
\begin{eqnarray}\label{11}
\mathbf{E}^{(+)}_{\alpha}&=&i\sqrt{\frac{2\hbar\omega_{\mathrm{cav}}}{\epsilon_0 L}}
\;a\big[\hat{\mathbf{r}}e_r\cos(\varphi-\varphi_0)\cos\beta_{\mathrm{cav}}(z-z_0)
\nonumber\\&&\mbox{}
+i\hat{\boldsymbol{\varphi}}e_\varphi\sin(\varphi-\varphi_0)\cos\beta_{\mathrm{cav}}(z-z_0)
\nonumber\\&&\mbox{}
+i\hat{\mathbf{z}}e_z\cos(\varphi-\varphi_0)\sin\beta_{\mathrm{cav}}(z-z_0)\big]e^{-i\omega_{\mathrm{cav}} t}.\qquad
\end{eqnarray}
Here, $a$ is the photon annihilation operator for the cavity mode. 
The notations $\hat{\mathbf{r}}=\hat{\mathbf{x}}\cos\varphi + \hat{\mathbf{y}}\sin\varphi$, 
$\hat{\boldsymbol{\varphi}}=-\hat{\mathbf{x}}\sin\varphi + \hat{\mathbf{y}}\cos\varphi$, and $\hat{\mathbf{z}}$
stand for the unit basis vectors of the cylindrical coordinate system, where $\hat{\mathbf{x}}$ and $\hat{\mathbf{y}}$ are the unit basis vectors of the Cartesian coordinate system for the fiber cross-section plane $xy$. The notations $e_r=e_r(r)$, $e_{\varphi}=e_{\varphi}(r)$, and $e_z=e_z(r)$ stand for
the cylindrical components of the profile function $\mathbf{e}^{(\omega_{\mathrm{cav}},+,+)}(r,\varphi)$ of the forward counterclockwise polarized guided mode 
at the cavity resonant frequency $\omega_{\mathrm{cav}}$ \cite{fiber books,cesium decay}. 
The parameter $z_0$ in Eqs.~\eqref{10} and \eqref{11} is given as $z_0=m\pi/2\beta_{\mathrm{cav}}$, where $m$ is an integer number and $\beta_{\mathrm{cav}}=\beta(\omega_{\mathrm{cav}})$ is the
guided mode propagation constant at the cavity resonant frequency. The angle $\varphi_0$ in Eq.~\eqref{11} determines the orientation of the polarization vector of the quasilinearly polarized nanofiber cavity mode. The prefactors in Eqs.~\eqref{10} and \eqref{11} are determined by the condition 
$2\epsilon_0\int_{-L/2}^{L/2} dz\int d^2\mathbf{r}\, n_{\mathrm{ref}}^2  \mathbf{E}^{(-)}_{\alpha}\mathbf{E}^{(+)}_{\alpha}=\hbar\omega_{\mathrm{cav}} a^\dagger a$.

In order to derive the coupling coefficients, we consider a two-level atom located at the position $(r,\varphi,z)$.
Let $\mathbf{d}=\langle +|\mathbf{D}|-\rangle$ be the matrix element of the electric dipole moment operator $\mathbf{D}$ for the atomic transition between the upper level
$|+\rangle$ and the lower level $|-\rangle$.
In general, the atomic dipole vector $\mathbf{d}$ of a realistic atom is a complex vector.
The interaction between the atom and the quantum guided cavity field in the dipole and rotating-wave approximations is described by the Hamiltonian 
\begin{equation}\label{12}
H_{AF}=-i\hbar(g a \sigma^\dagger-g^* a^\dagger \sigma),
\end{equation}
where the coupling coefficient $g$ is given, for quasicircular polarization, as
\begin{eqnarray}\label{13}
g&=&\sqrt{\frac{\omega_{\mathrm{cav}}}{\epsilon_0\hbar L}}
\mathbf{d}\cdot\big[\big(\hat{\mathbf{r}}e_r+l\hat{\boldsymbol{\varphi}}e_\varphi\big)\cos\beta_{\mathrm{cav}}(z-z_0)
\nonumber\\&&\mbox{}
+i\hat{\mathbf{z}}e_z\sin\beta_{\mathrm{cav}}(z-z_0)\big]e^{il\varphi}
\end{eqnarray}
and, for quasilinear polarization, as
\begin{eqnarray}\label{14}
g&=&\sqrt{\frac{2\omega_{\mathrm{cav}}}{\epsilon_0\hbar L}}
\mathbf{d}\cdot\big[\hat{\mathbf{r}}e_r\cos(\varphi-\varphi_0)\cos\beta_{\mathrm{cav}}(z-z_0)
\nonumber\\&&\mbox{}
+i\hat{\boldsymbol{\varphi}}e_\varphi\sin(\varphi-\varphi_0)\cos\beta_{\mathrm{cav}}(z-z_0)
\nonumber\\&&\mbox{}
+i\hat{\mathbf{z}}e_z\cos(\varphi-\varphi_0)\sin\beta_{\mathrm{cav}}(z-z_0)\big].
\end{eqnarray}
Note that $\Omega=2g$ is sometimes called the vacuum Rabi frequency. In Eq.~\eqref{12}, $\sigma=|-\rangle\langle+|$ and $\sigma^\dagger=|+\rangle\langle-|$
are the transition operators for a two-level atom.

We assume that the FBG mirrors do not reflect the radiation modes. This assumption is reasonable in the case where the distance $L$ between the FBG mirrors is large as compared to the fiber radius $a$
and to the wavelength $\lambda_0=2\pi/k_0$, with $k_0=\omega_0/c$ being the characteristic wave number of the atomic transitions. With this assumption, the mode functions of the radiation modes are not modified by the presence of the FBG mirrors. In other words, the radiation modes are not confined by the FBG cavity. 
In this sense, the physics of the FBG cavity is similar to that of one-dimensional cavities \cite{Cook,Feng}, and is different from that of planar Fabry-P\'{e}rot cavities \cite{CQED,Martini,Bjork,Dung}, where off-axis modes reduce the quantum electrodynamic (QED) effect of the cavity on spontaneous emission of the atom \cite{Bjork,Dung}. We also note that 
the guided field in the FBG cavity is confined not only in the axial direction between the mirrors but also in the fiber cross-section plane. In this sense, the physics of the FBG cavity is similar to that of curved Fabry-P\'{e}rot cavities, which are often used in experiments on cavity QED effects \cite{CQED,Rempe2000,Rempe2002,Kimble2004,Thompson92,Mabuchi,Kimble group,Rempe,Shimizu,McKeever,Maunz,Sauer}. An advantage of a FBG cavity based on a nanofiber is that the field in the guided modes can be confined to a small cross-section area whose size is comparable to the light wavelength \cite{nanofiber properties}. 
For example, for a nanofiber with radius of 250 nm, the effective cross-sectional mode area 
$A_{\mathrm{eff}}=(\int |\mathbf{e}^{(\mu)}|^2d\mathbf{r})^2/\int |\mathbf{e}^{(\mu)}|^4d\mathbf{r}$ 
of the quasicircularly polarized fundamental guided modes with the wavelength $\lambda=852$ nm is found to be $A_{\mathrm{eff}}\simeq 0.5$ $\mu\mathrm{m}^2$. The corresponding mode radius is found to be $r_{\mathrm{eff}}=\sqrt{A_{\mathrm{eff}}/\pi}\simeq 398$ nm. This value is much smaller than the typical values of 15 to 30 $\mu$m for the waists of the Fabry-P\'{e}rot cavity modes used in the experiments on cavity QED effects \cite{CQED,Rempe2000,Rempe2002,Kimble2004,Thompson92,Mabuchi,Kimble group,Rempe,Shimizu,McKeever,Maunz,Sauer}. 
The mode radius $r_{\mathrm{eff}}\simeq 398$ nm of a 250-nm-radius nanofiber is a few times smaller than
the mode waists between 1 and 2 $\mu$m of fiber Fabry-P\'{e}rot cavities \cite{Reichel2015}.

We drive the cavity by a classical light field propagating along the fiber in a guided mode 
$\mu_{p}=(\omega_{p}, f_{p}, \xi_{p})$. Let $P$ be the incident power. 
The pumping is described by the Hamiltonian
$\mathcal{H}_{p}=\hbar(\mathcal{E}_{p}a^\dagger+\mathcal{E}_{p}^*a)$,
where
\begin{equation}\label{16}
\mathcal{E}_{p}=\sqrt{\frac{\kappa}{2} \frac{P}{\hbar\omega_{p}}}
\end{equation}
is the cavity pumping rate. We assume that the FBG mirrors are lossless. Then,
the cavity damping rate is
\begin{equation}\label{17}
\kappa=\frac{(1-|R|^2)v_g}{|R|L},
\end{equation}
where $v_g=1/\beta'(\omega_{\mathrm{cav}})$ is the group velocity. 
The mean number $\bar{n}=\langle a^\dagger a\rangle$ of photons in the cavity without atoms is given by \cite{Walls} 
\begin{equation}\label{18}
\bar{n}=\frac{|\mathcal{E}_{p}|^2}{\kappa^2/4+(\omega_{p}-\omega_{\mathrm{cav}})^2}.
\end{equation}
The power $P^{\mathrm{(out)}}$ of the transmitted field is related to the mean intracavity photon number $\bar{n}$ via the formula \cite{Walls}
$P^{\mathrm{(out)}}/\hbar\omega_{p}=\kappa\bar{n}/2$.
The mean output photon number is 
\begin{equation}\label{19}
n^{\mathrm{(out)}}(t)=\frac{1}{2}\int_{-\infty}^t\kappa\bar{n}(t')dt'.
\end{equation}
The reflected field results from the interference of the field that is directly reflected at the incoupling mirror and the field that issues from inside the cavity. 
The power $P^{\mathrm{(ref)}}$ of the reflected field depends on not only the mean intracavity photon number $\bar{n}$ but also the mean intracavity photon amplitude $\langle a\rangle$ via the formula
$P^{\mathrm{(ref)}}/\hbar\omega_{p}=\kappa\bar{n}/2+2|\mathcal{E}_{p}|^2/\kappa+i\langle a^\dagger \mathcal{E}_{p} -a \mathcal{E}_{p}^* \rangle $. 
Information about the mean intracavity photon amplitude $\langle a\rangle$ can be obtained from the power $P^{\mathrm{(ref)}}$ of the reflected field. When the cavity is at exact resonance and the atom is not present in the cavity or does not interact with the cavity field, the reflected field is zero.
For the switching operation in the present paper, we are interested in the effect of the atom on the intracavity field and the transmitted field.

The cooperativity parameter is defined as $\eta=4|g|^2/\gamma_0\kappa$.
We note that the cooperativity parameter $\eta$ does not depend on the cavity length $L$.

As known \cite{CQED}, the regimes of the interaction between an atom and a quantum field in an optical cavity are determined by the atom-field coupling coefficient $g$, the cavity damping rate $\kappa$, and the atomic decay rate $\gamma_0$. 
In order to achieve the strong-coupling regime, it is desirable to have a cavity with a large $g$ and a small $\kappa$. For the same set of the values of the parameters $g$ and $\kappa$, a nanofiber cavity 
and a microcavity used in the experiments on cavity QED effects \cite{CQED,Rempe2000,Rempe2002,Kimble2004,Thompson92,Mabuchi,Kimble group,Rempe,Shimizu,McKeever,Maunz,Sauer}
can be very different from each other in the cavity length $L$, the free spectral range $\Delta_{\mathrm{FSR}}$, and the cavity finesse $\mathcal{F}$. Indeed, the coupling coefficient $g$ depends on the effective mode volume $V_{\mathrm{eff}}=A_{\mathrm{eff}}L$ via the formula $g\propto 1/\sqrt{V_{\mathrm{eff}}}$. In order for $g$ to be large, $V_{\mathrm{eff}}$ must be small. In a nanofiber-based cavity, the nanofiber and the FBG mirrors confine the guided field in the transverse plane and the longitudinal direction, respectively. 
As already mentioned, the mode matching between the cavity field and the guided modes of the nanofiber is very good in the experiments \cite{milling,Ding11,Wuttke12,Nayak13,Sadgrove13,Nayak14,Yalla14,Aoki15}.
Due to the tight confinement of the guided field of the nanofiber, the effective cross-sectional mode area $A_{\mathrm{eff}}$ is small. Therefore, $V_{\mathrm{eff}}$ can be small and consequently $g$ can be large even when $L$ is large. 
Since the cavity length $L$ can be large, the free spectral range $\Delta_{\mathrm{FSR}}=\pi v_g/L$ can be small. 
This is desirable in order to tune two longitudinal modes of the cavity into resonance with two atomic transitions with the transition frequency difference of about $9.2$ GHz. Furthermore, the cavity damping rate $\kappa=\Delta_{\mathrm{FSR}}/\mathcal{F}=\pi v_g/\mathcal{F}L$ can be small even when the cavity finesse $\mathcal{F}=\Delta_{\mathrm{FSR}}/\kappa=\pi |R|/(1-|R|^2)$ is moderate.
Since the cavity finesse $\mathcal{F}$ can be moderate, the mirror reflectivity $|R|^2$ does not have to be very high and, therefore,
the mirror transmittivity $|T|^2$ can be significant. This is good in order to couple the intracavity field to the outside world. 
The above features of the nanofiber-based cavity have been employed to demonstrate the strong-coupling regime and the vacuum Rabi splitting in an all-fiber cavity system with a single trapped atom \cite{Aoki15}. The evanescent-wave nature of the mode functions of the guided field leads to the tight confinement in the fiber transverse plane and to the efficient coupling. Another feature of the nanofiber-based cavity is that the cavity guided field can be transmitted over long distances for communication purposes.

\section{Dressed states of the coupled atom-cavity system}
\label{sec:dressed}

The eigenstates of the Hamiltonian $\mathcal{H}_0$ are called the dressed states of the atom-cavity system.
We use the notation $|j,n_1,n_2\rangle$ for bare states, where $j$ is the atomic level index and $n_1$ and $n_2$ are the numbers of photons in cavity modes 1 and 2, respectively.
In terms of the bare states $|1,n_1,n_2\rangle$, $|2,n_1-1,n_2\rangle$, $|3,n_1-1,n_2\rangle$, and $|4,n_1-1,n_2-1\rangle$, which are the basis for the manifold $(n_1,n_2)$,
the Hamiltonian $\mathcal{H}_0$ can be presented as a matrix consisting of the blocks 
\begin{widetext}
\begin{equation}\label{20}
\mathcal{H}_0^{(n_1n_2)}=\hbar
\begin{pmatrix} -\Delta_{\mathrm{cav}_1}n_1-\Delta_{\mathrm{cav}_2}n_2&0&ig_1^*\sqrt{n_1}&0\\
0&\delta_2-\Delta_{\mathrm{cav}_1}n_1-\Delta_{\mathrm{cav}_2}n_2&-\Omega_c^*/2&ig_2^*\sqrt{n_2}\\
-ig_1\sqrt{n_1}&-\Omega_c/2&\delta_3-\Delta_{\mathrm{cav}_1}n_1-\Delta_{\mathrm{cav}_2}n_2&0\\
0&-ig_2\sqrt{n_2}&0&\delta_4-\Delta_{\mathrm{cav}_1}n_1-\Delta_{\mathrm{cav}_2}n_2
\end{pmatrix}.
\end{equation}
\end{widetext}
Consequently, the dressed states belonging to the manifold $(n_1,n_2)$ are superpositions of the bare states $|1,n_1,n_2\rangle$, $|2,n_1-1,n_2\rangle$, $|3,n_1-1,n_2\rangle$, and $|4,n_1-1,n_2-1\rangle$. 
The dressed states are denoted as $|\psi^{(n_1n_2)}_j\rangle$, where $j=1$ for $n_1=0$ and $n_2\geq0$, $j=1,2,3$ for $n_1\geq1$ and $n_2=0$, and $j=1,2,3,4$ for $n_1,n_2\geq 1$. The energies of these eigenstates are denoted as $\hbar\epsilon^{(n_1n_2)}_j$. 
The explicit expressions for the dressed states and their energies have been derived for the case where
the different atomic transitions $|3\rangle\leftrightarrow|1\rangle$ and $|4\rangle\leftrightarrow|2\rangle$ interact with the same cavity mode \cite{Rebic02a}.
We extend the results of Ref.~ \cite{Rebic02a} for the case where the different atomic transitions $|3\rangle\leftrightarrow|1\rangle$ and $|4\rangle\leftrightarrow|2\rangle$ interact with
the different cavity modes.

\subsubsection{Manifolds $(0,n_2)$ with $n_2\geq0$}

Each of the manifolds $(0,n_2)$ where $n_2\geq0$ contains only one state, namely
$|\psi^{(0,n_2)}_1\rangle=|1,0,n_2\rangle$.
The energy of this state is $\hbar\epsilon^{(0,n_2)}_1=-\hbar\Delta_{\mathrm{cav}_2}n_2$.

\subsubsection{Manifolds $(n_1,0)$ with $n_1\geq1$}

Each of the manifolds $(n_1,0)$ where $n_1\geq1$ contains three dressed states.
The energies of these eigenstates can be written as $\hbar\epsilon^{(n_1,0)}_j=\hbar\tilde{\epsilon}_j-\hbar\Delta_{\mathrm{cav}_1}n_1$, where $\tilde{\epsilon}_j$ with $j=1,2,3$ are the roots of the cubic equation $x^3+u_2x^2+u_1x+u_0=0$. Here, we have introduced the notations
\begin{equation}\label{21}
\begin{split}
u_0&=|g_1|^2n_1 \delta_2 ,\\
u_1&=-|g_1|^2n_1  - \frac{|\Omega_c|^2}{4} + \delta_2 \delta_3,\\
u_2&= - \delta_2 - \delta_3.
\end{split}
\end{equation}
Using the explicit expressions for the roots \cite{Abramovich}, we find
\begin{equation}\label{22}
\begin{split}
\tilde{\epsilon}_1&=-\frac{1}{3}u_2+(s_1+s_2),\\
\tilde{\epsilon}_2&=-\frac{1}{3}u_2-\frac{1}{2}(s_1+s_2)+i\frac{\sqrt3}{2}(s_1-s_2),\\
\tilde{\epsilon}_3&=-\frac{1}{3}u_2-\frac{1}{2}(s_1+s_2)-i\frac{\sqrt3}{2}(s_1-s_2),
\end{split}
\end{equation}
where
\begin{equation}\label{23}
\begin{split}
s_1&=(r+\sqrt{r^2+q^3})^{1/3},\\
s_2&=-\frac{q}{(r+\sqrt{r^2+q^3})^{1/3}},
\end{split}
\end{equation}
with
\begin{equation}\label{24}
\begin{split}
r&=\frac{1}{6}(u_1u_2-3u_0)-\frac{1}{27}u_2^3,\\
q&=\frac{1}{3}u_1-\frac{1}{9}u_2^2.
\end{split}
\end{equation}

The expressions for the corresponding eigenstates are given as
\begin{equation}\label{25}
\begin{split}
|\psi^{(n_1,0)}_j\rangle&=A^{(n_1,0)}_j|1,n_1,0\rangle+B^{(n_1,0)}_j|2,n_1-1,0\rangle\\
&\quad +C^{(n_1,0)}_j|3,n_1-1,0\rangle,
\end{split}
\end{equation}
where
\begin{equation}\label{26}
\begin{split}
A^{(n_1,0)}_j&=(1+|U_1|^2+|U_2|^2)^{-1/2},\\
B^{(n_1,0)}_j&=U_1A^{(n_1,0)}_j,\\
C^{(n_1,0)}_j&=U_2A^{(n_1,0)}_j,
\end{split}
\end{equation}
with 
\begin{equation}\label{27}
\begin{split}
U_1&=-\frac{ig_1\sqrt{n_1}}{\Omega_c/2}\bigg[1-\frac{\tilde{\epsilon}_j(\tilde{\epsilon}_j-\delta_3)}{|g_1|^2n_1}\bigg],\\
U_2&=\frac{\tilde{\epsilon}_j}{ig_1^*\sqrt{n_1}}.
\end{split}
\end{equation}

In the particular case where the two-photon detuning for the transition $|1\rangle\leftrightarrow|2\rangle$ is $\delta_2=0$, we find the eigenvalues 
\begin{equation}\label{28}
\begin{split}
\tilde{\epsilon}_0&=0,\\
\tilde{\epsilon}_{\pm}&=\frac{1}{2}\delta_3\pm\sqrt{|g_1|^2n_1+\frac{1}{4}|\Omega_c|^2+\frac{1}{4}\delta_3^2}.
\end{split}
\end{equation}
The expressions for the corresponding eigenstates are
\begin{equation}\label{29}
\begin{split}
|\psi^{(n_1,0)}_0\rangle&=A^{(n_1,0)}_0|1,n_1,0\rangle+B^{(n_1,0)}_0|2,n_1-1,0\rangle,  \\
|\psi^{(n_1,0)}_{\pm}\rangle&=A^{(n_1,0)}_{\pm}|1,n_1,0\rangle+B^{(n_1,0)}_{\pm}|2,n_1-1,0\rangle\\
&\quad +C^{(n_1,0)}_{\pm}|3,n_1-1,0\rangle,
\end{split}
\end{equation}
where 
\begin{equation}\label{30}
\begin{split}
A^{(n_1,0)}_0&=\frac{\Omega_c/2}{\sqrt{|g_1|^2n_1+|\Omega_c|^2/4}},\\
B^{(n_1,0)}_0&=-\frac{ig_1\sqrt{n_1}}{\sqrt{|g_1|^2n_1+|\Omega_c|^2/4}}, 
\end{split}
\end{equation}
and
\begin{equation}\label{31}
\begin{split}
A^{(n_1,0)}_{\pm}&=\frac{ig_1^*\sqrt{n_1}}{\sqrt{|g_1|^2n_1+|\Omega_c|^2/4+|\tilde{\epsilon}_{\pm}|^2}},\\
B^{(n_1,0)}_{\pm}&=-\frac{\Omega_c^*/2}{\sqrt{|g_1|^2n_1+|\Omega_c|^2/4+|\tilde{\epsilon}_{\pm}|^2}},\\
C^{(n_1,0)}_{\pm}&=\frac{\tilde{\epsilon}_{\pm}}{\sqrt{|g_1|^2n_1+|\Omega_c|^2/4+|\tilde{\epsilon}_{\pm}|^2}}.
\end{split}
\end{equation}
The eigenstates $|\psi^{(n_1,0)}_0\rangle$ do not contain any upper levels and their energies are not shifted
by the atom-field interaction. These states are the dark states.

\subsubsection{Manifolds $(n_1,n_2)$ with $n_1,n_2\geq1$}

Each of the manifolds $(n_1,n_2)$ where $n_1,n_2\geq1$ contains four dressed states.
The energies of these eigenstates 
can be written as $\hbar\epsilon^{(n_1n_2)}_j=\hbar\tilde{\epsilon}_j-\hbar\Delta_{\mathrm{cav}_1}n_1-\hbar\Delta_{\mathrm{cav}_2}n_2$, where $\tilde{\epsilon}_j$ 
with $j=1,2,3,4$ are the roots of the quartic equation $x^4+b_3x^3+b_2x^2+b_1x+b_0=0$. Here, we have introduced the notations
\begin{align}\label{32}
b_0&=|g_1|^2n_1(|g_2|^2n_2-\delta_2 \delta_4),\nonumber\\
b_1&=|g_1|^2n_1(\delta_2+\delta_4) + |g_2|^2n_2\delta_3 + \frac{|\Omega_c|^2}{4}\delta_4-\delta_2 \delta_3 \delta_4,\nonumber\\
b_2&= - |g_1|^2n_1 - |g_2|^2n_2 -\frac{|\Omega_c|^2}{4}+\delta_2 (\delta_3 + \delta_4) + \delta_3 \delta_4,\nonumber\\
b_3&=-\delta_2 - \delta_3 - \delta_4.
\end{align}
Using the explicit expressions for the roots \cite{Abramovich}, we find
\begin{equation}\label{33}
\begin{split}
\tilde{\epsilon}_{1,2}&=-\frac{1}{4}b_3+\frac{1}{2}R\pm \frac{1}{2}P,\\
\tilde{\epsilon}_{3,4}&=-\frac{1}{4}b_3-\frac{1}{2}R\pm \frac{1}{2}Q,
\end{split}
\end{equation}
where
\begin{equation}\label{34}
\begin{split}
R&=\sqrt{\frac{1}{4}b_3^2-\frac{2}{3}b_2 +S},\\
P&=\sqrt{\frac{1}{2}b_3^2-\frac{4}{3}b_2-S+W},\\
Q&=\sqrt{\frac{1}{2}b_3^2-\frac{4}{3}b_2-S-W}.
\end{split}
\end{equation}
The parameter $S$ is defined as
\begin{equation}\label{35}
S=(X+\sqrt{X^2+Y^3})^{1/3}-\frac{Y}{(X+\sqrt{X^2+Y^3})^{1/3}},
\end{equation}
where
\begin{equation}\label{37}
\begin{split}
X&=\frac{1}{2}b_3^2b_0-\frac{1}{6}b_3b_2b_1+\frac{1}{27}b_2^3-\frac{4}{3}b_2b_0+\frac{1}{2}b_1^2,\\
Y&=\frac{1}{3}b_3b_1-\frac{1}{9}b_2^2-\frac{4}{3}b_0.
\end{split}
\end{equation}
The parameter $W$ is defined as
\begin{equation}\label{36}
W=\begin{cases}
(b_3b_2-2b_1-b_3^3/4)/R   &\text{if $R\not=0$},\\
2[(S+b_2/3)^2-4b_0]^{1/2} &\text{if $R=0$}.
\end{cases}
\end{equation}

The expressions for the corresponding eigenstates are
\begin{eqnarray}\label{38}
|\psi^{(n_1n_2)}_j\rangle&=&A^{(n_1n_2)}_j|1,n_1,n_2\rangle+B^{(n_1n_2)}_j|2,n_1-1,n_2\rangle \nonumber\\
&&\mbox{} +C^{(n_1n_2)}_j|3,n_1-1,n_2\rangle \nonumber\\
&&\mbox{} +D^{(n_1n_2)}_j|4,n_1-1,n_2-1\rangle,
\end{eqnarray}
where
\begin{gather}\label{39}
\frac{A^{(n_1n_2)}_j}{D^{(n_1n_2)}_j}=V_1 ,\quad
\frac{B^{(n_1n_2)}_j}{D^{(n_1n_2)}_j}=V_2 ,\quad
\frac{C^{(n_1n_2)}_j}{D^{(n_1n_2)}_j}=V_3 ,\nonumber\\
D^{(n_1n_2)}_j=(1+|V_1|^2+|V_2|^2+|V_3|^2)^{-1/2},
\end{gather}
with
\begin{eqnarray}\label{40}
V_1&=&-\frac{g_1^*g_2^*\sqrt{n_1n_2}}{\tilde{\epsilon}_j\Omega_c^*/2}\bigg[1-\frac{(\tilde{\epsilon}_j-\delta_2) (\tilde{\epsilon}_j-\delta_4)}{|g_2|^2n_2}\bigg],
\nonumber\\
V_2&=&i\frac{\tilde{\epsilon}_j - \delta_4}{g_2\sqrt{n_2}},
\nonumber\\
V_3&=&i\frac{g_2^*\sqrt{n_2}}{\Omega_c^*/2}\bigg[1-\frac{(\tilde{\epsilon}_j-\delta_2) (\tilde{\epsilon}_j-\delta_4)}{|g_2|^2n_2}\bigg].
\end{eqnarray}

The above analytical results remain valid for the eigenstates and eigenvalues of the non-Hermitian Hamiltonian $\mathcal{H}_0+\mathcal{H}_{\mathrm{damp}}$
if we replace the cavity resonant frequencies $\omega_{\mathrm{cav}_1}$ and $\omega_{\mathrm{cav}_2}$ by $\omega_{\mathrm{cav}_1}-i\kappa_1/2$ and $\omega_{\mathrm{cav}_2}-i\kappa_2/2$, respectively, 
and replace the atomic level frequencies $\omega_2$, $\omega_3$, and $\omega_4$ by $\omega_2-i\gamma_{2\mathrm{deph}}/2$, $\omega_3-i\gamma_3/2$, and $\omega_4-i\gamma_4/2$, respectively.

\begin{figure}[tbh]
\begin{center}
\includegraphics{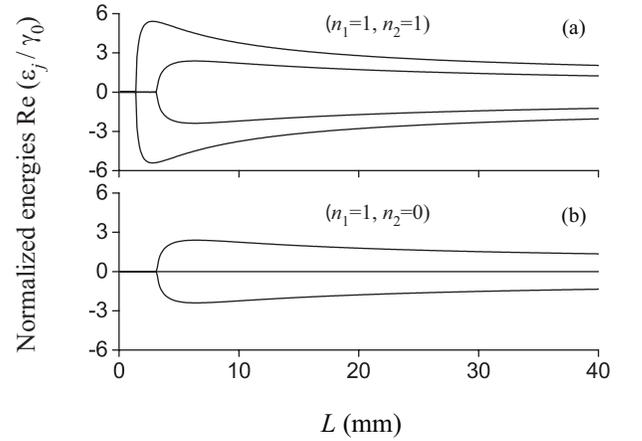}
\end{center}
\caption{Normalized energies $\text{Re}(\epsilon_j/\gamma_0)$ of the dressed states in the manifolds $(n_1=1,n_2=1)$ (a) and $(n_1=1,n_2=0)$ (b)
as functions of the cavity length $L$ for $|R|^2=0.99$. The atomic levels are specified in the caption of Fig.~\ref{fig1}.
The intensity of the control field is $I_c=5$ $\mathrm{mW/cm}^2$. 
The external control field is linearly polarized along the $y$ axis, while the cavity guided modes are quasilinearly polarized along the $x$ axis.
The detunings are $\Delta_{\mathrm{cav}_1}=\Delta_{\mathrm{cav}_2}=\Delta=\delta=0$. The fiber radius is $a=250$ nm. The distance from the atom to the fiber surface is $r-a=200$ nm. 
The axial position of the atom coincides with an antinode of the radial component of the cavity field. 
}
\label{fig2}
\end{figure} 

We use the above analytical results to calculate the eigenstates and eigenvalues of the Hamiltonian $\mathcal{H}_0+\mathcal{H}_{\mathrm{damp}}$. 
We use the energy levels of the $D_2$ line of atomic cesium specified in the caption of Fig.~\ref{fig1}. 
The fiber radius is $a=250$ nm and the distance from the atom to the fiber surface is $r-a=200$ nm \cite{Vetsch10}.
The external control field is linearly polarized along the $y$ axis, while the cavity guided modes are quasilinearly polarized along the $x$ axis.
The axial position of the atom coincides with an antinode of the radial component of the cavity field. 
The coupling coefficients $g_1$ and $g_2$ are calculated from Eq.~\eqref{14} for $z=z_0$ and $\varphi=\varphi_0$.
The decay rates $\gamma_3\simeq 2\pi\times 5.34$ MHz and $\gamma_4\simeq 2\pi\times 5.44$ MHz are obtained by using the results of Ref. \cite{cesium decay}. 
Due to the presence of the nanofiber, the values of $\gamma_3$ and $\gamma_4$ are slightly larger than the natural linewidth $\gamma_0\simeq 2\pi\times5.2$ MHz of the $D_2$ line of atomic cesium  \cite{Steck02,coolingbook}. The lower-level decoherence rate is assumed to be $\gamma_{2\mathrm{deph}}/2=2\pi\times 50$ kHz. 
This value is comparable to the experimental value of about $2\pi\times 32$ kHz, measured for the magnetic-field-insensitive transition $M=0\leftrightarrow M'=0$ in the Vienna experiment \cite{Mitsch14a}. 
The fields are at exact resonance with the corresponding atomic transitions. 

We plot the normalized energies $\text{Re}(\epsilon_j/\gamma_0)$ of the dressed states in the manifolds $(n_1=1,n_2=1)$ and $(n_1=1,n_2=0)$ as functions of the cavity length $L$ 
in Fig.~\ref{fig2} for the case where the mirror reflectivity is $|R|^2=0.99$.
The figure shows that the level splitting occurs when the cavity length $L$ is large enough. 
We observe from the lower part of Fig.~\ref{fig2} that, among the eigenstates in the manifold
$(n_1=1,n_2=0)$, there is always a state whose energy is not shifted by the atom-field interaction. 
This state is the dark state $|\psi^{(1,0)}_0\rangle$, which does not contain any upper levels and therefore is a long-lived state.
We observe from the upper part of Fig.~\ref{fig2} that 
the energies of the eigenstates in the manifold $(n_1=1,n_2=1)$ are shifted by the atom-field interaction except for the region of small $L$ where the cavity damping rates and, consequently, the cavity mode linewidths are much larger than the atom-field coupling coefficients. 
The eigenstates $|\psi^{(1,1)}_j\rangle$ with $j=1,\dots,4$ in the manifold $(n_1=1,n_2=1)$ contain the upper levels and hence are the bright states.
The generation of the bright states $|\psi^{(1,1)}_j\rangle$ and the coupling between them by the cavity pump fields include the possibility of simultaneous absorption of a photon in mode 1 and a photon in mode 2. In addition, the bright states
are shorter lived than the dark state $|\psi^{(1,0)}_0\rangle$ of the manifold $(n_1=1,n_2=0)$. 
Therefore, the presence of a photon in cavity mode 2 may reduce the possibility of having
a photon in cavity mode 1.

\section{Switch for cavity mode 1}
\label{sec:option1}

In this section, we show that we can switch the field in cavity mode 1 by using the field in cavity mode 2.
The switch is realized by using the field in cavity mode 2 as a gate for the conventional EIT scheme that is based on the atomic levels $|1\rangle$, $|2\rangle$, and $|3\rangle$
with the field in cavity mode 1 as the probe field and the external field $\mathcal{E}_c$ as the control field.
A similar all-optical switch has been experimentally demonstrated for a small laser-cooled ensemble of atoms inside a hollow fiber with running-wave gate and probe light fields \cite{Bajcsy09}. 

\begin{figure}[tbh]
\begin{center}
\includegraphics{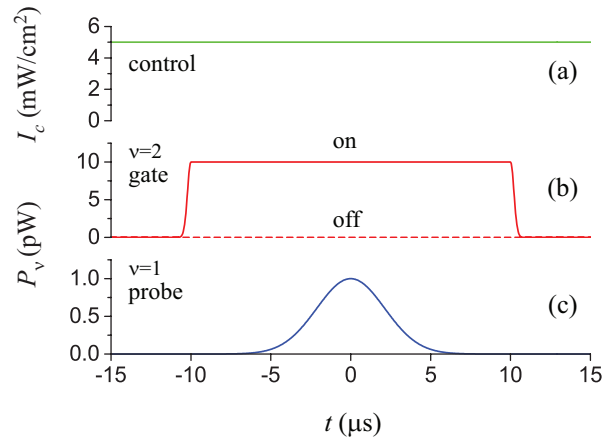}
\end{center}
\caption{(Color online)
Switch for cavity mode 1 by using the field in cavity mode 2.
The time dependencies of the intensity $I_c$ of the control field (a), the driving field power $P_2$ for the gate mode (b),  
and the driving field power $P_1$ for the probe mode (c) are plotted.
The external control field is linearly polarized along the $y$ axis, while the guided driving fields for the cavity modes are quasilinearly polarized along the $x$ axis.
The intensity of the control field is constant in the whole process and is $I_c=5 \text{ mW/cm}^2$. 
The driving field for cavity mode 2 is on (red solid line) or off (red dashed line). 
In the case of gate on, the power of the driving field for mode 2 is constant and equal to $P_2^{(\mathrm{max})}=10$ pW in the time interval $(-T_2,T_2)$, where $T_2=10$ $\mu$s. 
The ascending and descending parts of the driving pulse for cavity mode 2 are of Gaussian shape, with a characteristic width of $0.5$ $\mu$s. 
The driving field for cavity mode 1 is a Gaussian pulse with the peak power $P_1^{\mathrm{(max)}}=1$ pW and the full width at half maximum $T_1=5$ $\mu$s. 
}
\label{fig3}
\end{figure} 

We apply the control field $\mathcal{E}_c$ and the driving pulses $\mathcal{E}_1$ and $\mathcal{E}_2$ for cavity modes 1 and 2 in a time sequence shown in Fig.~\ref{fig3}.
The external control field $\mathcal{E}_c$ is linearly polarized along the $y$ axis, while the guided driving fields $\mathcal{E}_1$ and $\mathcal{E}_2$ for the cavity modes are quasilinearly polarized along the $x$ axis. The intensity of the control field $\mathcal{E}_c$ is constant for the whole process and is chosen to be $I_c=5 \text{ mW/cm}^2$. 
The guided driving field $\mathcal{E}_2$ for cavity mode 2, which is used as the gate field, is either on (red solid line) or off (red dashed line). 
In the case of gate on, the power of $\mathcal{E}_2$ is $P_2^{(\mathrm{max})}=10$ pW in the time interval $(-T_2,T_2)$, where $T_2=10$ $\mu$s.
The ascending and descending parts of $\mathcal{E}_2$ are of Gaussian shape, with a full width at half maximum of $0.5$ $\mu$s. 
The guided driving field for cavity mode 1, which is used as the probe field, is a Gaussian pulse with the peak power $P_1^{\mathrm{(max)}}=1$~pW and the full width at half maximum $T_1=5$ $\mu$s. 
Note that the gate driving pulse contains about 880 photons on average, while the probe driving pulse contains about $23$ photons on average.
The Rabi frequency of the control field is $\Omega_c/2\pi\simeq5.4$ MHz $\simeq 1.03 \gamma_0$.

\begin{figure}[tbh]
\begin{center}
\includegraphics{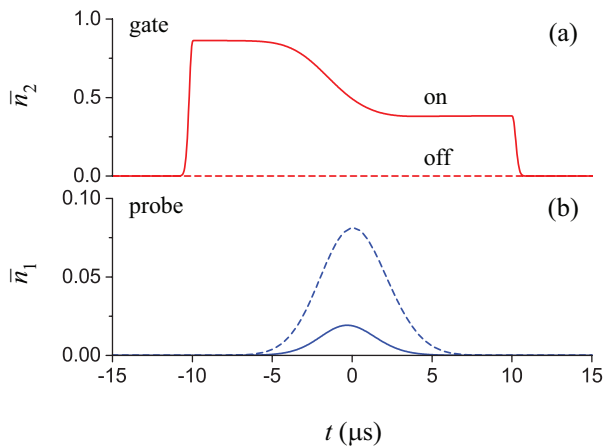}
\end{center}
\caption{(Color online)
Switching of the field in cavity mode 1 by using the field in cavity mode 2.
The time dependencies of the mean numbers $\bar{n}_2$ (a)  and $\bar{n}_1$ (b) of photons in the cavity gate and probe modes, respectively, are plotted. The solid and dashed curves correspond to the cases where the field in gate mode 2 is on and off, respectively. The atomic levels are specified in the caption of Fig.~\ref{fig1}, while
the time sequence, the durations, and the magnitudes of the control and cavity driving fields are as in Fig.~\ref{fig3}.
The fiber radius is $a=250$ nm. The distance from the atom to the fiber surface is $r-a=200$ nm. The axial position of the atom coincides with an antinode of the radial component of the cavity field.  
The reflectivity of the cavity mirrors is $|R|^2=0.99$. The cavity length is $L=20$ mm.
The detunings are $\Delta_{\mathrm{cav}_1}=\Delta_{\mathrm{cav}_2}=\Delta=\delta=0$.
}
\label{fig4}
\end{figure} 

We illustrate in Fig.~\ref{fig4} the switching of the field in cavity mode 1 by using the field in cavity mode 2.
The time dependencies of the mean numbers $\bar{n}_2$  and $\bar{n}_1$ of photons in the cavity gate and probe modes, respectively, are plotted.  
The atomic levels are specified in the caption of Fig.~\ref{fig1}, while the time sequence, the durations, and the magnitudes of the control and cavity driving fields are as in Fig.~\ref{fig3}.
The fiber radius is $a=250$ nm. The distance from the atom to the fiber surface is $r-a=200$ nm. The axial position of the atom coincides with an antinode of the radial component of the cavity field. 
The reflectivity of the cavity mirrors is $|R|^2=0.99$. The cavity length is $L=20$ mm.
The corresponding value of the free spectral range is $\Delta_{\mathrm{FSR}}/2\pi\simeq 4.9$ GHz.
The cavity finesse is $\mathcal{F}_1=\mathcal{F}_2=\mathcal{F}\simeq313$. 
The cavity damping rates are $\kappa_1/2\pi=\kappa_2/2\pi=\kappa/2\pi\simeq15.8$ MHz.
The coupling parameters are calculated from Eq.~\eqref{14} for $z=z_0$ and $\varphi=\varphi_0$ and are found to be $|g_1|/2\pi\simeq9.4$ MHz and $|g_2|/2\pi\simeq14.6$ MHz. 
The cooperativity parameters for modes 1 and 2 are $\eta_1\simeq4.3$ and $\eta_2\simeq10.3$, respectively.
These values indicate that the strong-coupling regime ($\eta>1$) can be realized even though the cavity is long and the cavity finesse is moderate \cite{cavityspon}.
Comparison between the solid (gate-on) and dashed (gate-off) curves of Fig.~\ref{fig4}(b) shows that the turn on and turn off of the field in cavity mode 2 significantly affect the magnitude of the field in cavity mode 1. The switching action can occur even though the mean numbers $\bar{n}_2$ and $\bar{n}_1$ of photons 
in the gate and probe modes, respectively, are smaller than 1. The ratio between the peak values of the mean number $\bar{n}_1$ of photons in probe mode 1 in the gate-off and gate-on cases [see the blue dashed and blue solid lines in Fig.~\ref{fig4}(b)] is equal to about $4.29$.

\begin{figure}[tbh]
\begin{center}
\includegraphics{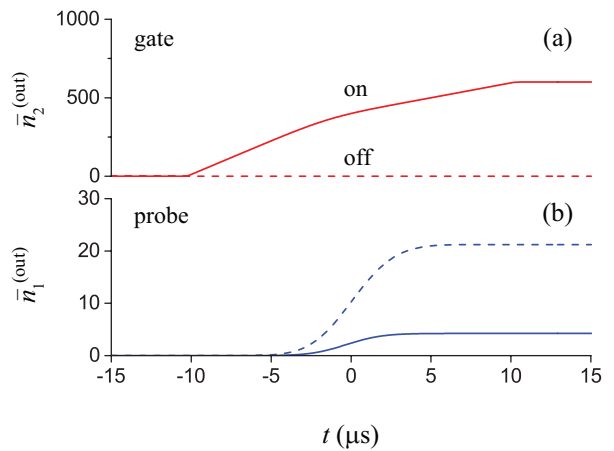}
\end{center}
\caption{(Color online)
Time dependencies of the mean output photon numbers $\bar{n}_2^{(\mathrm{out})}$ (a) and $\bar{n}_1^{(\mathrm{out})}$ (b) in the cases where cavity mode field 2 is on (solid lines) or off (dashed lines). The parameters used are as for Figs.~\ref{fig3} and \ref{fig4}.
}
\label{fig5}
\end{figure} 

In order to get deep insight into the switching operation, we plot in Fig.~\ref{fig5} the time dependencies of the mean output photon numbers $\bar{n}_2^{(\mathrm{out})}$  and $\bar{n}_1^{(\mathrm{out})}$ in the cases where cavity mode field 2 is on (solid lines) or off (dashed lines). 
The ratio $\xi_1$ between the long-time limiting values of the mean number $\bar{n}_1^{(\mathrm{out})}$ of photons in the output of probe mode 1 in the switch-off and switch-on cases 
[see the blue dashed and blue solid lines in Fig.~\ref{fig5}(b)], is equal to about 5. The switching contrast, given as $(\xi_1-1)/(\xi_1+1)$, is about 67\%.
It is clear that the transmission reduction factor $\xi_1$ for the scheme considered here cannot exceed 
the transmission reduction factor $(1+\eta_1)^2$ for cavity mode 1 interacting with a two-level atom at exact resonance \cite{Suzuki2011a,Walls}. Here, 
$\eta_1\simeq4.3$ is, as already stated, the cooperativity parameter for mode 1 in the case considered. 
We note that the mean output photon number $\bar{n}_2^{(\mathrm{out})}$ for gate mode 2 shown in Fig.~\ref{fig5}(a) is rather large.
The reason is that the gate-mode driving field $\mathcal{E}_2$ is kept constant for a time interval $(-T_2,T_2)$, where $T_2=10$ $\mu$s. 
This time interval is larger than the full width at half maximum $T_1=5$ $\mu$s of the probe pulse $\mathcal{E}_1$.  
We can reduce $\bar{n}_2^{(\mathrm{out})}$ by reducing $T_1$ and $T_2$. 
However, we should not reduce $T_1$ and $T_2$ too much. If $T_1$ is too small, the EIT for the field in probe mode 1 in the case of gate off will be significantly reduced and so is the switching contrast.

As already mentioned in the discussion below Eq.~\eqref{19},
information about the mean intracavity photon amplitude can be obtained from the power of the reflected field. 
However, for the purpose of the switching operation, we are interested in the effect of the atom on the mean intracavity photon number and the mean number of transmitted photons.

\begin{figure}[tbh]
\begin{center}
\includegraphics{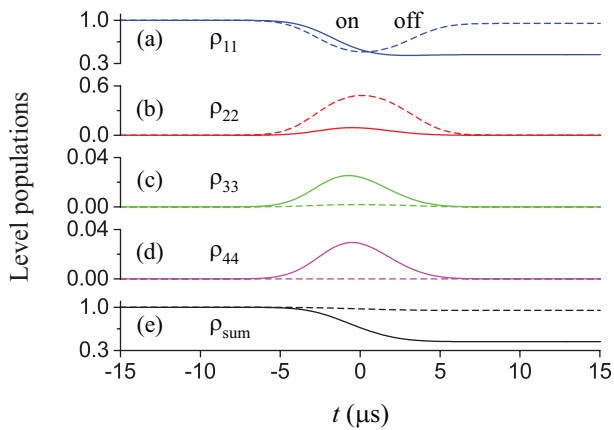}
\end{center}
\caption{(Color online)
Time dependencies of the populations $\rho_{jj}$ of the atomic energy levels $|j\rangle$ with $j=1$ (a), 2 (b), 3 (c), and 4 (d) and the sum population $\rho_{\mathrm{sum}}$
of the four working levels (e) in the cases where cavity mode field 2 is on (solid lines) or off (dashed lines).
The parameters used are as for Figs.~\ref{fig3} and \ref{fig4}.
}
\label{fig6}
\end{figure} 

We plot in Fig.~\ref{fig6} the time dependencies of the populations $\rho_{jj}$ of the atomic energy levels $|j\rangle$ with $j=1,2,3,4$ and the sum population $\rho_{\mathrm{sum}}$
of the four working levels. We observe from the dashed lines of Fig.~\ref{fig6} that, in the case where the gate is off,
the interaction of the atom with the probe and control fields leads to a coherent adiabatic population transfer between the lower levels $|1\rangle$ and $|2\rangle$. In this process, 
the atom adiabatically follows its dark state, and the excitation of the upper level $|3\rangle$ is very weak. This is the situation of EIT in its broad meaning \cite{review}. 
Strictly speaking, the case of the dashed lines of Fig.~\ref{fig6}, where the coherent adiabatic population transfer between the lower levels is significant,
is different from the conventional EIT process, where the atom practically remains in its ground state $|1\rangle$ \cite{review}.
We note that, when we reduce the peak power $P_1^{\mathrm{(max)}}$ of the driving pulse for probe mode 1 to $0.1$ pW as done in the case of Fig.~\ref{fig12}(a),
the atom in the absence of the gate field will practically remain in the ground state $|1\rangle$. 

A close inspection of the curves of Fig.~\ref{fig6}(a) shows that, in the time region $t<0$, 
the population $\rho_{11}$ in the case of gate on (solid curve) is larger than in the case of gate off (dashed curve). 
This means that the simultaneous absorption of a gate photon and a probe photon by the atom is not the only mechanism for the suppression of $\bar{n}_1$. 
Another important mechanism for the switching action is the photon blockade \cite{Imamoglu97,Werner99,Rebic99,Gheri99,Greentree00,Rebic02a,Rebic02b,Bajcsy13} caused by the presence of a photon in cavity mode 2. We observe from Fig.~\ref{fig6}(e) that the total population $\rho_{\mathrm{sum}}=\sum_{j=1}^4\rho_{jj}$ of the four working levels $|j\rangle$ with $j=1,\dots,4$ is not
conserved in the evolution process. The reason is that we have $\gamma_3>\gamma_{31}+\gamma_{32}$. This formula is a consequence of the fact that, in the case of atomic cesium,
the population of the upper level $|3\rangle$ can decay not only to the lower levels $|1\rangle$ and $|2\rangle$ but 
also to some other lower levels which are outside of the working level configuration and, therefore, are not shown in Fig.~\ref{fig1}. The deviation of $\rho_{\mathrm{sum}}$ from the unity is substantial in the case of gate on, where the coupled atom-cavity system is excited to bright states, 
but is negligible in the case of gate off, where the system adiabatically follows the dark state under the EIT conditions. 

\begin{figure}[tbh]
\begin{center}
\includegraphics{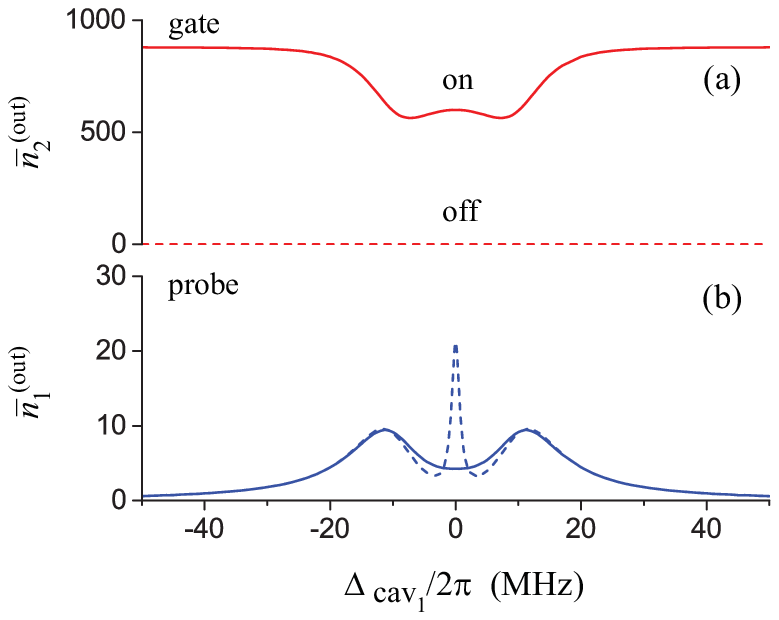}
\end{center}
\caption{(Color online)
Dependencies of the mean output photon numbers $\bar{n}_2^{(\mathrm{out})}$ (a) and $\bar{n}_1^{(\mathrm{out})}$ (b) on the detuning $\Delta_{\mathrm{cav}_1}$ of the probe driving pulse $\mathcal{E}_1$ with respect to the resonant frequency of cavity mode 1 in the cases where cavity mode field 2 is on (solid lines) or off (dashed lines). The outputs are integrated from the beginning of the interaction process to the time $t=20$ $\mu$s after the peak time of the probe driving pulse $\mathcal{E}_1$. Other parameters are as for Figs.~\ref{fig3} and \ref{fig4}. 
In particular, the resonance condition $\Delta_{\mathrm{cav}_2}=0$ for the pump for mode 2 and the two-photon resonance condition $\delta_2=0$ for the atom-cavity interaction are maintained.
}
\label{fig7}
\end{figure} 

We plot in Fig.~\ref{fig7} the dependencies of the mean output photon numbers $\bar{n}_2^{(\mathrm{out})}$ and $\bar{n}_1^{(\mathrm{out})}$ on the detuning $\Delta_{\mathrm{cav}_1}$ of the probe driving pulse $\mathcal{E}_1$ with respect to the resonant frequency of cavity mode 1 in the cases where cavity mode field 2 is on (solid lines) or off (dashed lines). 
The resonance condition $\Delta_{\mathrm{cav}_2}=0$ for the pump for mode 2 and the two-photon resonance condition $\delta_2=0$ for the atom-cavity interaction are maintained. The blue dashed curve in Fig.~\ref{fig7}(b) shows clearly the existence of a narrow cavity EIT peak in the frequency dependence of the output of cavity mode field 1 in the case where cavity mode field 2 is off. Under the conditions $\gamma_3\gamma_{2\mathrm{deph}}\ll|\Omega_c|^2 \ll |g_1|^2 $, the width $\kappa_{\mathrm{EIT}}$ of the central cavity EIT window can be estimated as $\kappa_{\mathrm{EIT}}=\gamma_{2\mathrm{deph}}+\kappa |\Omega_c|^2/4|g_1|^2$ \cite{Dantan12}. 
The blue solid curve in Fig.~\ref{fig7}(b) shows that there is no cavity EIT peak in the case where cavity mode field 2 is on. Thus, the excitation of the field in cavity mode 2 destroys the EIT for the field in cavity mode 1. We note that the side peaks in Fig.~\ref{fig7}(b) are the signature of the vacuum Rabi splitting of the cavity transmission spectrum of the field in probe mode 1 \cite{Suzuki2011a,Muecke2010}.

\begin{figure}[tbh]
\begin{center}
\includegraphics{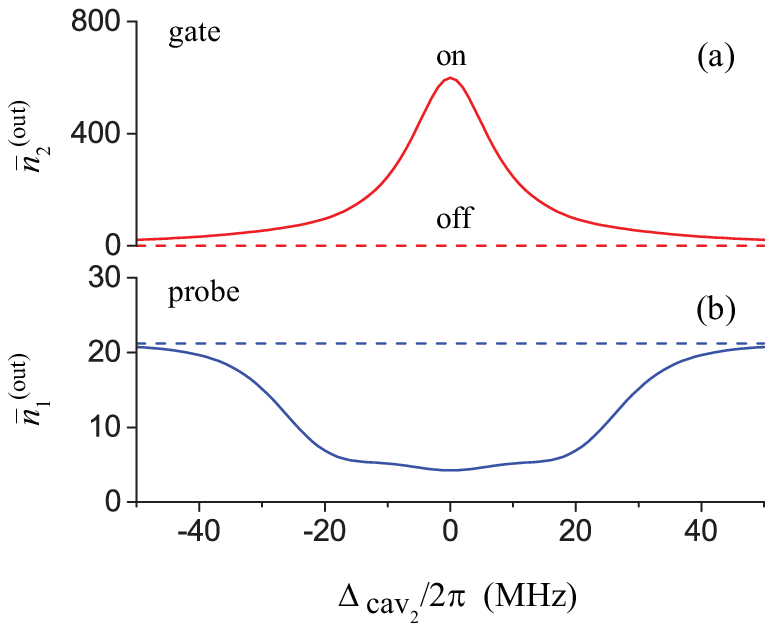}
\end{center}
\caption{(Color online)
Dependencies of the mean output photon numbers $\bar{n}_2^{(\mathrm{out})}$ (a) and $\bar{n}_1^{(\mathrm{out})}$ (b) on the detuning $\Delta_{\mathrm{cav}_2}$ of the driving
gate pulse $\mathcal{E}_2$ with respect to the resonant frequency of cavity mode 2 in the cases where the field in this mode is on (solid lines) or off (dashed lines). The outputs are integrated from the beginning of the interaction process to the time $t=20$ $\mu$s after the peak time of the probe driving pulse $\mathcal{E}_1$. Other parameters are as for Figs.~\ref{fig3} and \ref{fig4}. 
In particular, the resonance condition $\Delta_{\mathrm{cav}_1}=0$ for the pump for mode 1 and the two-photon resonance condition $\delta_2=0$ for the atom-cavity interaction are maintained.
}
\label{fig8}
\end{figure} 

We plot in Fig.~\ref{fig8} the dependencies of the mean output photon numbers $\bar{n}_2^{(\mathrm{out})}$ and $\bar{n}_1^{(\mathrm{out})}$ on the detuning $\Delta_{\mathrm{cav}_2}$ of the gate driving pulse $\mathcal{E}_2$ with respect to the resonant frequency of cavity mode 2 in the cases where the field in this mode is on (solid lines) or off (dashed lines). 
The resonance condition $\Delta_{\mathrm{cav}_1}=0$ for the pump for mode 1 and the two-photon resonance condition $\delta_2=0$ for the atom-cavity interaction are maintained.
Figure~\ref{fig8}(b) shows that the cavity EIT for mode 1 is suppressed by the field in mode 2 in a wide region of the detuning $\Delta_{\mathrm{cav}_2}$. The size of this region is determined by the vacuum Rabi frequency of cavity gate mode 2, that is, by the coupling coefficient $|g_2|/2\pi\simeq14.6$ MHz.

\begin{figure}[tbh]
\begin{center}
\includegraphics{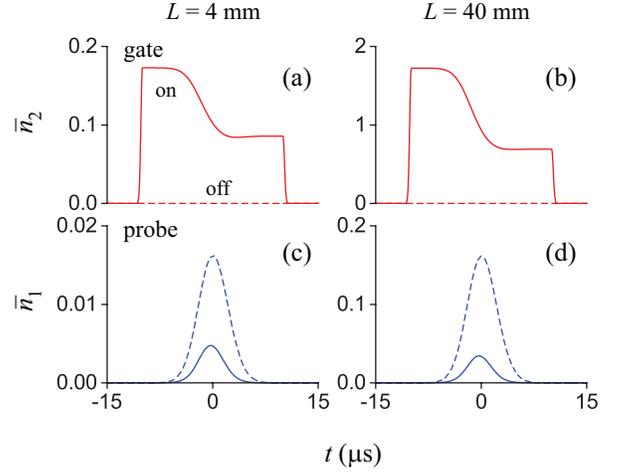}
\end{center}
\caption{(Color online)
Switching of the field in mode 1 by using the field in mode 2 for 
the cavity length $L=4$ mm (left column) and $L=40$ mm (right column).
The solid and dashed curves correspond to the cases where the field in gate mode 2 is on and off, respectively.
Other parameters are as for Figs.~\ref{fig3} and \ref{fig4}.
}
\label{fig9}
\end{figure} 

\begin{figure}[tbh]
\begin{center}
\includegraphics{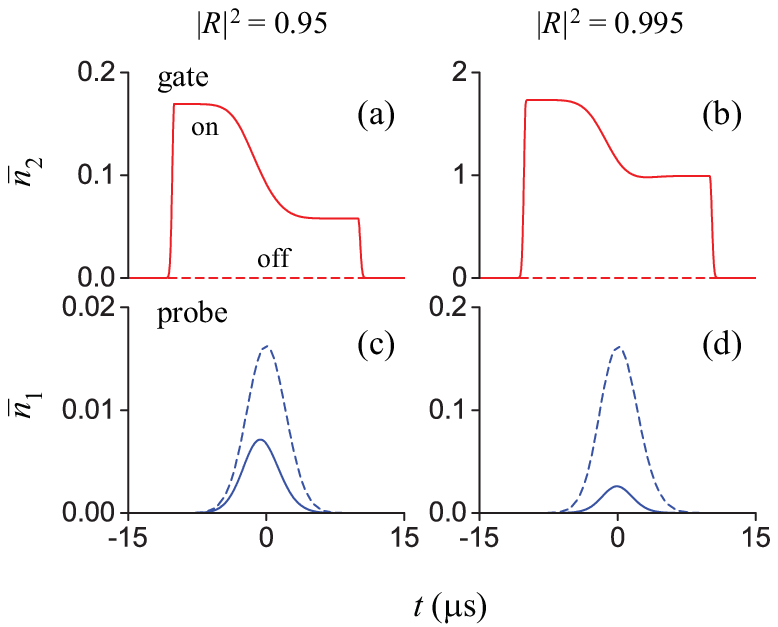}
\end{center}
\caption{(Color online)
Switching of the field in mode 1 by using the field in mode 2 for 
the FBG mirror reflectivity $|R|^2=0.95$ (left column) and $|R|^2=0.995$ (right column).
The solid and dashed curves correspond to the cases where the field in gate mode 2 is on and off, respectively.
Other parameters are as for Figs.~\ref{fig3} and \ref{fig4}.
}
\label{fig10}
\end{figure} 

We illustrate in Figs.~\ref{fig9} and \ref{fig10} the switching operations for two different values of the cavity length $L$ and two different values of the FBG mirror reflectivity $|R|^2$, respectively. Comparison between the left and right columns of the figures shows that the mean cavity-mode photon numbers $\bar{n}_2$ and $\bar{n}_1$ in the right column, where $L=40$ mm in the case of Fig.~\ref{fig9} and $|R|^2=0.995$ ($\mathcal{F}\simeq 627$) in the case of Fig.~\ref{fig10}, are larger than those in the left column, where $L=4$ mm in the case of Fig.~\ref{fig9} and $|R|^2=0.95$ ($\mathcal{F}\simeq 61$) in the case of Fig.~\ref{fig10}. The reason is that the values of the damping rate $\kappa$ in the case of the right column is smaller than that in the case of the left column. We observe from Figs.~\ref{fig9} and \ref{fig10} that the switching contrast, that is, 
the suppression of the mean number $\bar{n}_1$ of photons in the probe mode, increases with increasing cavity length or increasing mirror reflectivity. We observe that the effect of the cavity length $L$ on the switching contrast is weak, while the effect of the reflectivity $|R|^2$ on the switching contrast is strong. These features are the consequence of the fact that the suppression of the probe field is mainly determined by the cooperativity parameter 
$\eta_1=4|g_1|^2/\gamma_0\kappa$, which depends on $|R|^2$ but does not depend on $L$.

\begin{figure}[tbh]
\begin{center}
\includegraphics{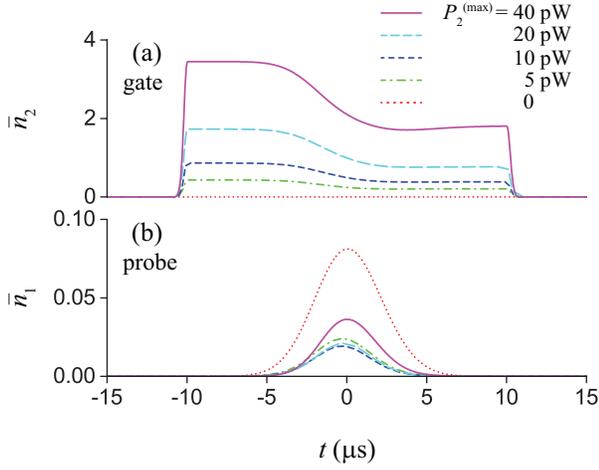}
\end{center}
\caption{(Color online)
Switching of the field in mode 1 by using the field in mode 2 for 
different values of $P_2^{(\mathrm{max})}$. 
The solid and dashed curves correspond to the cases where the field in gate mode 2 is on and off, respectively.
Other parameters are as for Figs.~\ref{fig3} and \ref{fig4}.
}
\label{fig11}
\end{figure}

Figure \ref{fig11} illustrates the switching operations for different values of the peak power $P_2^{(\mathrm{max})}$ of the driving pulse for cavity gate mode 2.  
We observe from the figure that, when we increase the power $P_2^{(\mathrm{max})}$, 
the suppression of the mean photon number $\bar{n}_1^{(\mathrm{out})}$ of cavity probe mode 1 first
increases and then decreases. This feature indicates that the interaction process in the cavity is nonlinear.

\begin{figure}[tbh]
\begin{center}
\includegraphics{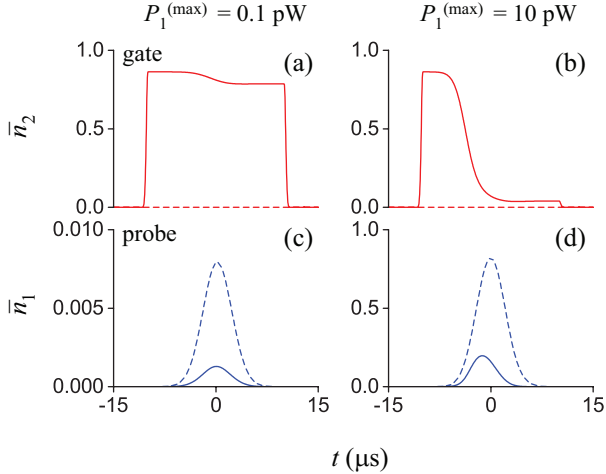}
\end{center}
\caption{(Color online)
Switching of the field in mode 1 by using the field in mode 2 for 
different values of $P_1^{(\mathrm{max})}$. The solid and dashed curves correspond to the cases where the field in gate mode 2 is on and off, respectively. Other parameters are as for Figs.~\ref{fig3} and \ref{fig4}.
}
\label{fig12}
\end{figure}

Figure \ref{fig12} illustrates the switching operations for two different values of the peak power $P_1^{(\mathrm{max})}$ of the pump pulse for cavity probe mode 1.  
We observe from the figure that the suppression of the mean photon number $\bar{n}_1^{(\mathrm{out})}$ of cavity probe mode 1
decreases with increasing power $P_1^{(\mathrm{max})}$. 
This decrease of the suppression factor is a result of the power broadening effect.

\begin{figure}[tbh]
\begin{center}
\includegraphics{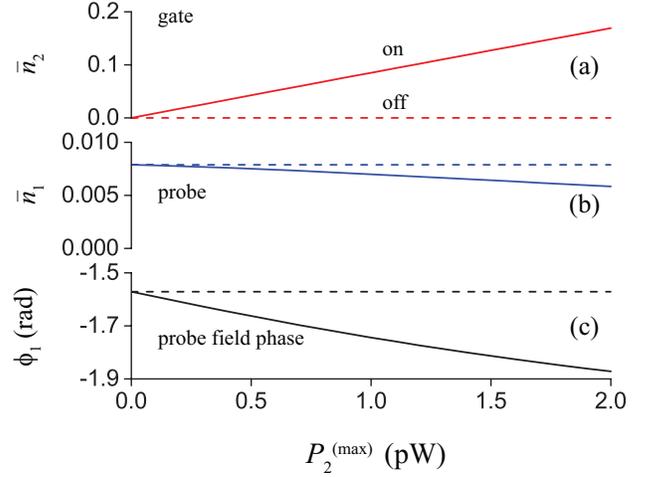}
\end{center}
\caption{(Color online)  Operation of the switch in the dispersive regime. 
Cavity gate mode 2 and the corresponding driving field are equally detuned from the atomic transition $|4\rangle\leftrightarrow|2\rangle$ by a detuning $\Delta/2\pi=-100$ MHz
with $\Delta_{\mathrm{cav}_2}=0$. The peak power of the driving field for cavity probe mode 1 is $P_1^{\mathrm{(max)}}=0.1$ pW. 
Other parameters are as for Figs.~\ref{fig3} and \ref{fig4}.
The mean photon number $\bar{n}_2$ of mode 2, the mean photon number $\bar{n}_1$ of mode 1, 
and the phase $\phi_1$ of the complex amplitude $\langle a_1\rangle$ of mode 1
at the probe pulse peak time $t=0$ are plotted as functions of $P_2^{(\mathrm{max})}$ (solid lines).
For comparison, the results for the case of gate off are shown by the dashed curves. 
}
\label{fig13}
\end{figure}

When the driving field for gate mode 2 is far detuned from the atomic transition $|4\rangle\leftrightarrow |2\rangle$, the main effect of the cavity field in gate mode 2 on the atom is to produce light shifts of the energy levels $|4\rangle$ and $|2\rangle$. If the light shifts are comparable to or greater than the width of the cavity EIT window,
the transmission of the field in probe mode 1 is suppressed, as the cavity is switched off resonance by the presence of the field in gate mode 2. An EIT-based light switch using ion Coulomb crystals in an optical cavity with a gate field in the dispersive regime has recently been demonstrated \cite{Albert11,Dantan12}. If the light shift of the level $|2\rangle$ is within the cavity EIT window for probe mode 1, the EIT is not destroyed. However, due to the steep dispersion of the atom-field interaction in the vicinity of the EIT window, the phase shift of the field in probe mode 1 can be significant. This phase shift is proportional to the number of photons in gate mode 2. We illustrate the dispersive regime of the operation of the switch in Fig.~\ref{fig13}, where  cavity gate mode 2 and the driving field for this mode are equally detuned from the atomic transition $|4\rangle\leftrightarrow|2\rangle$ by a detuning 
$\Delta/2\pi=-100$ MHz with $\Delta_{\mathrm{cav}_2}=0$. In this figure, we plot the mean photon number $\bar{n}_2$ of mode 2, the mean photon number $\bar{n}_1$ of mode 1, and the phase $\phi_1$ of the complex amplitude $\langle a_1\rangle$ of mode 1 at the probe pulse peak time $t=0$ as functions of the peak power $P_2^{(\mathrm{max})}$ of the gate driving pulse. 
The peak power of the driving field for cavity probe mode 1 is $P_1^{\mathrm{(max)}}=0.1$ pW. Other parameters are as for Figs.~\ref{fig3} and \ref{fig4}.
Figure~\ref{fig13} shows that, when we increase $P_2^{(\mathrm{max})}$ from 0 to 2 pW, at the probe pulse peak time $t=0$, the mean photon number $\bar{n}_2$ of mode 2 increases linearly from 0 to $\simeq 0.17$, the mean photon number $\bar{n}_1$ of mode 1 decreases slightly from $\simeq 0.008$ to $\simeq 0.006$, and the phase $\phi_1$ of the complex amplitude $\langle a_1\rangle$ of mode 1 decreases linearly from $\simeq -1.57$ to $\simeq -1.87$ rad. Comparison between Figs.~\ref{fig13}(a) and \ref{fig13}(c) shows that the phase shift per intracavity gate photon is $\simeq 1.76$ rad/photon.

\section{Switch for cavity mode 2}
\label{sec:option2}

In this section, we show that we can switch the field in cavity mode 2 by manipulating the field in cavity mode 1.
In other words, we can use the fields in cavity modes 1 and 2 as the gate and probe fields, respectively.
For this purpose, we adopt the scheme demonstrated experimentally by Chen \textit{et al.} for an ensemble of atoms inside a high-finesse ($\mathcal{F}\simeq 6.3\times 10^4$) optical cavity \cite{Chen13}.

\begin{figure}[tbh]
\begin{center}
\includegraphics{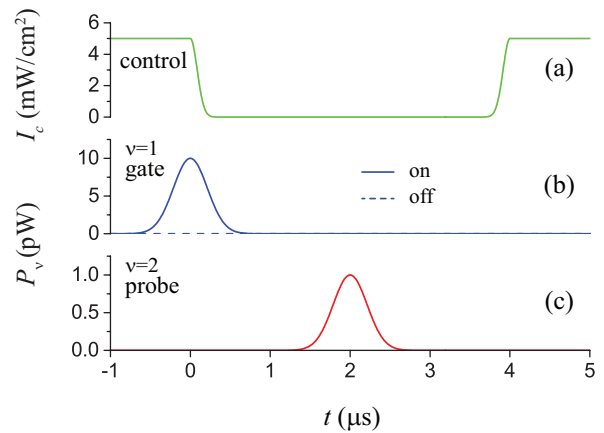}
\end{center}
\caption{(Color online)
Switch for cavity mode 2 by storing a photon of cavity mode 1 in the atomic lower level $|2\rangle$.
The time dependencies of the control field intensity $I_c$ (a), the driving field power $P_1$ for the gate mode (b), and the driving field power $P_2$ for the probe mode (c) are plotted.
The external control field is linearly polarized along the $y$ axis, while the guided driving fields for the cavity modes are quasilinearly polarized along the $x$ axis.
The intensity of the control field is $I_c=5 \text{ mW/cm}^2$ in the time regions $t<0$ and $t>4$ $\mu$s. 
The control field is ramped down at $t=0$ and is re-applied to reach the previous intensity value $I_c=5 \text{ mW/cm}^2$ at $t=4$ $\mu$s.
The descending and ascending parts of the control field are of Gaussian shape, with a characteristic width of $0.2$ $\mu$s. 
The driving field for cavity mode 1 is either on (blue solid line) or off (blue dashed line). It has
a Gaussian shape with the full width at half maximum $T_1=0.5$ $\mu$s and the peak power $P_1^{\mathrm{(max)}}=10$ pW.
The driving field for cavity mode 2 is a Gaussian pulse with the full width at half maximum $T_2=0.5$ $\mu$s, the peak time $t_2=2$ $\mu$s, and the peak power $P_2^{\mathrm{(max)}}=1$ pW.
}
\label{fig14}
\end{figure}

The time sequence for the application of the pulses is shown in Fig.~\ref{fig14}.
First, we apply an external field $\mathcal{E}_c$ to create the EIT conditions for the field in cavity mode 1 and then
send in a weak guided field $\mathcal{E}_{p_1}$ to excite cavity mode 1. 
Around the arrival time of the peak of the gate pulse $\mathcal{E}_{p_1}$ [around the time $t=0$ in Fig.~\ref{fig14}(b)], 
we ramp down the control field $\mathcal{E}_c$ to store a gate photon in the atomic lower level $|2\rangle$ 
\cite{Fleischhauer2000,Liu2001,Phillips2001}. At a later time [around the time $t=4$ $\mu$s in Fig.~\ref{fig14}(a)], we retreat this photon by reapplying the control field $\mathcal{E}_c$.
In between the storage and retrieval stages [around the time $t=2$ $\mu$s in Fig.~\ref{fig14}(c)], we send in a weak guided field $\mathcal{E}_{p_2}$ to excite cavity mode 2. 
The population of the atomic state $|2\rangle$ created by the stored gate photon reduces the transmission of the probe pulse $\mathcal{E}_{p_2}$ through the cavity.
The magnitude of the reduction factor in the case of a two-level atom at exact resonance is given by the quantity $1/(1+\eta_2)^2$, 
where $\eta_2=4|g_2|^2/\gamma_0\kappa_2\simeq10.3$ is the cooperativity parameter for mode 2 \cite{Walls, Suzuki2011a}.  

As illustrated in Fig.~\ref{fig14}, we use the control field $\mathcal{E}_c$ with a constant intensity $I_c=5 \text{ mW/cm}^2$ in the time regions $t<0$ and $t>4$ $\mu$s.
The driving field $\mathcal{E}_1$ for cavity mode 1, which is used as the gate field, is either on (blue solid line) or off (blue dashed line). 
When is turned on, the driving pulse $\mathcal{E}_1$ has a Gaussian shape with the full width at half maximum $T_1=0.5$ $\mu$s and the peak power $P_1^{\mathrm{(max)}}=10$ pW.
The driving field for cavity mode 2, which is used as the probe field, is a Gaussian pulse with the full width at half maximum $T_2=0.5$ $\mu$s, the peak time $t_2=2$ $\mu$s, and the peak power $P_2^{\mathrm{(max)}}=1$ pW. The external control field is linearly polarized along the $y$ axis, while the guided driving fields for the cavity modes are quasilinearly polarized along the $x$ axis.
The Rabi frequency of the control field is $\Omega_c/2\pi\simeq5.4$ MHz $\simeq 1.03 \gamma_0$.
Note that the gate driving pulse contains about 23 photons on average, while the probe driving pulse contains about $2.3$ photons on average.

\begin{figure}[tbh]
\begin{center}
\includegraphics{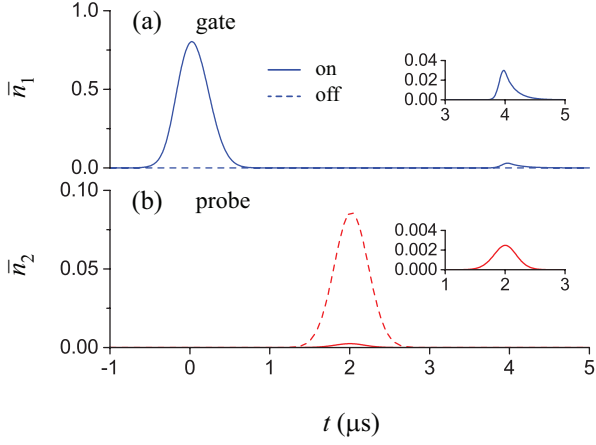}
\end{center}
\caption{(Color online)
Switching of the field in cavity mode 2 by using the field in cavity mode 1.
The time dependencies of the mean numbers $\bar{n}_1$ (a)  and $\bar{n}_2$ (b) of photons in the cavity gate and probe modes, respectively, are plotted.  
The solid and dashed curves correspond to the cases where the field in gate mode 1 is on and off, respectively.
The atomic levels are specified in the caption of Fig.~\ref{fig1}, while the time sequence, the durations, and the magnitudes of the control and driving fields are as in Fig.~\ref{fig14}.
The fiber radius is $a=250$ nm. The distance from the atom to the fiber surface is $r-a=200$ nm. The axial position of the atom coincides with an antinode of the radial component of the cavity field.  
The reflectivity of the cavity mirrors is $|R|^2=0.99$. The cavity length is $L=20$ mm.
The detunings are $\Delta_{\mathrm{cav}_1}=\Delta_{\mathrm{cav}_2}=\Delta=\delta=0$.
}
\label{fig15}
\end{figure} 

We illustrate in Fig.~\ref{fig15} the switching of the field in cavity mode 2 by using the field in cavity mode 1.
The time dependencies of the mean numbers $\bar{n}_1$  and $\bar{n}_2$ of photons in the cavity gate and probe modes, respectively, are plotted.  
The atomic levels are specified in the caption of Fig.~\ref{fig1}, while the time sequence, the durations, and the magnitudes of the control and driving fields are as in Fig.~\ref{fig14}.
As in the previous section, the fiber radius is $a=250$ nm, the distance from the atom to the fiber surface is $r-a=200$ nm, the axial position of the atom coincides with an antinode of the radial component of the cavity field, the reflectivity of the cavity mirrors is $|R|^2=0.99$, and the cavity length is $L=20$ mm.
For these parameters, we obtain, as already mentioned in the previous section, 
the free spectral range $\Delta_{\mathrm{FSR}}/2\pi\simeq 4.9$ GHz,
the cavity finesse $\mathcal{F}_1=\mathcal{F}_2=\mathcal{F}\simeq313$, 
the cavity damping rates $\kappa_1/2\pi=\kappa_1/2\pi=\kappa/2\pi\simeq15.8$ MHz,
the coupling parameters $|g_1|/2\pi\simeq 9.4$ MHz and $|g_2|/2\pi\simeq14.6$ MHz,
and the cooperativity parameters $\eta_1\simeq 4.3$ and $\eta_2\simeq 10.3$.
Comparison between the solid (gate-on) and dashed (gate-off) curves of Fig.~\ref{fig15}(b) shows that the turn on and turn off of the field in cavity mode 1 significantly affect the magnitude of the field in cavity mode 2. The switching action can occur even though the mean numbers $\bar{n}_1$ and $\bar{n}_2$ of photons 
in the gate and probe modes, respectively, are small (less than one).
The ratio between the peak values of the mean number $\bar{n}_2$ of photons in probe mode 2 in the switch-off and switch-on cases [see the red solid and red dashed lines in Fig.~\ref{fig15}(b)] is equal to about 35. The inset in Fig.~\ref{fig15}(a) shows the retrieval of the gate photon stored in the atomic lower level $|2\rangle$. 

\begin{figure}[tbh]
\begin{center}
\includegraphics{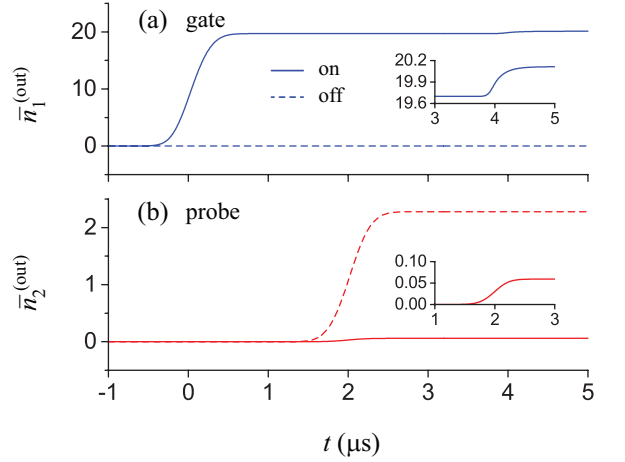}
\end{center}
\caption{(Color online)
Time dependencies of the mean output photon numbers $\bar{n}_1^{(\mathrm{out})}$ (a) and $\bar{n}_2^{(\mathrm{out})}$ (b) in the cases where cavity mode field 1 is on (solid lines) or off (dashed lines). The parameters used are as for Figs.~\ref{fig14} and \ref{fig15}.
}
\label{fig16}
\end{figure} 

We plot in Fig.~\ref{fig16} the time dependencies of the mean output photon numbers $\bar{n}_1^{(\mathrm{out})}$  and $\bar{n}_2^{(\mathrm{out})}$ in the cases where cavity mode field 1 is on (solid lines) or off (dashed lines). The ratio $\xi_2$ between the long-time limiting values of the mean number $\bar{n}_2^{(\mathrm{out})}$ of photons in the output of probe mode 2 in the switch-off and switch-on cases [see the red dashed and red solid lines in Fig.~\ref{fig16}(b)], is equal to about $38.5$. The switching contrast, given by $(\xi_2-1)/(\xi_2+1)$, is about 95\%. 
The significant reduction of $\bar{n}_2^{(\mathrm{out})}$ is due to the interaction between the atom and cavity field mode 2 in the strong-coupling regime. We note that the magnitude of the reduction factor is not in  perfect agreement with the semiclassical estimate $(1+\eta_2)^2\simeq 128$ \cite{Suzuki2011a,Walls}, where $\eta_2=4|g_2|^2/\gamma_0\kappa_2\simeq10.3$ is the cooperativity parameter for mode 2. 
One reason is that the population of the level $|1\rangle$ is not completely transferred to the level $|2\rangle$. Another reason is that 
a nonzero dephasing rate $\gamma_{2\mathrm{deph}}/2=2\pi\times 50$ kHz is used in our numerical calculations. In addition, 
the semiclassical approximation used in deriving the transmission reduction factor $1/(1+\eta_2)^2$ is not well justified for the parameters used.

The inset in Fig.~\ref{fig16}(a) shows that the mean output photon number $\bar{n}_1^{(\mathrm{out})}$ 
varies quickly from $19.7$ to $20.1$ around the time $t=4$ $\mu$s, when the control field is re-applied. Taking into account the fact that the nanofiber cavity is a two-sided cavity, we find that the mean number of retreated photons is equal to about $0.8$.

\begin{figure}[tbh]
\begin{center}
\includegraphics{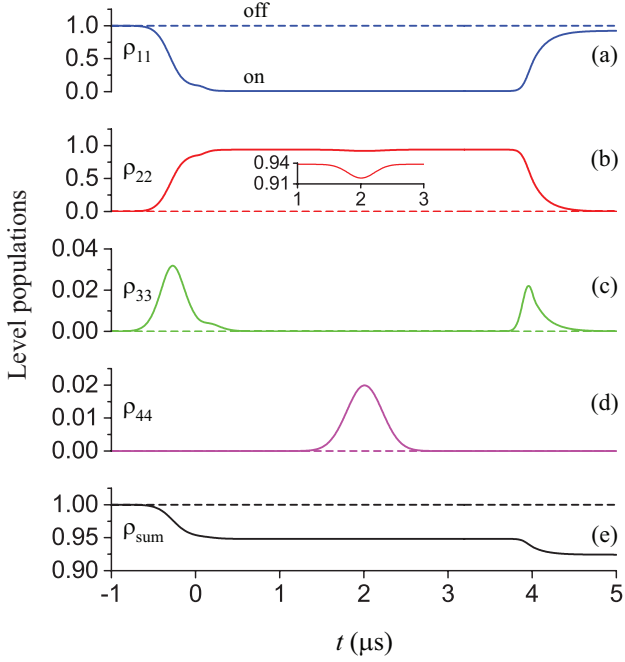}
\end{center}
\caption{(Color online)
Time dependencies of the populations $\rho_{jj}$ of the atomic energy levels $|j\rangle$ with $j=1$ (a), 2 (b), 3 (c), and 4 (d) and the sum population $\rho_{\mathrm{sum}}$
of the four working levels (e) in the cases where cavity mode field 1 is on (solid lines) or off (dashed lines).
The parameters used are as for Figs.~\ref{fig14} and \ref{fig15}.
}
\label{fig17}
\end{figure}

We plot in Fig.~\ref{fig17} the time dependencies of the populations $\rho_{jj}$ of the atomic energy levels $|j\rangle$ with $j=1,2,3,4$ and the sum population $\rho_{\mathrm{sum}}$
of the four working levels. The solid curves of Figs.~\ref{fig17}(a)--\ref{fig17}(c) show that, in the region around the time $t=0$, when the storage is performed, the population of the atom
is almost completely transferred from the level $|1\rangle$ to the level $|2\rangle$. This means that a gate photon is stored in the population of the level $|2\rangle$ of the atom with
a high probability. The solid curves of Figs.~\ref{fig17}(b) and \ref{fig17}(d) show that, in the region around the time $t=2$ $\mu$s, 
when the probe pulse  is sent in, a fraction of the population of the atom
moves from the level $|2\rangle$ to the level $|4\rangle$ and then returns to the level $|2\rangle$. The peak magnitude of the transferred population is about $0.02$. This quantity is significant even though the corresponding peak value of the mean photon number $\bar{n}_2$ is very small (about $0.0025$)  [see the inset of Fig.~\ref{fig15}(b)]. 
The solid curves of Figs.~\ref{fig17}(a)--\ref{fig17}(c) show that, in the region around the time $t=4$ $\mu$s, when the retrieval is performed, the population of the level $|2\rangle$ is transferred back to the level $|1\rangle$. We observe from Fig.~\ref{fig17}(e) that the total population $\rho_{\mathrm{sum}}=\sum_{j=1}^4\rho_{jj}$ of the four working levels $|j\rangle$ with $j=1,\dots,4$ is not
conserved in the case where the gate is on. The reason is that, as already mentioned in the previous section, we have $\gamma_3>\gamma_{31}+\gamma_{32}$. 

\begin{figure}[tbh]
\begin{center}
\includegraphics{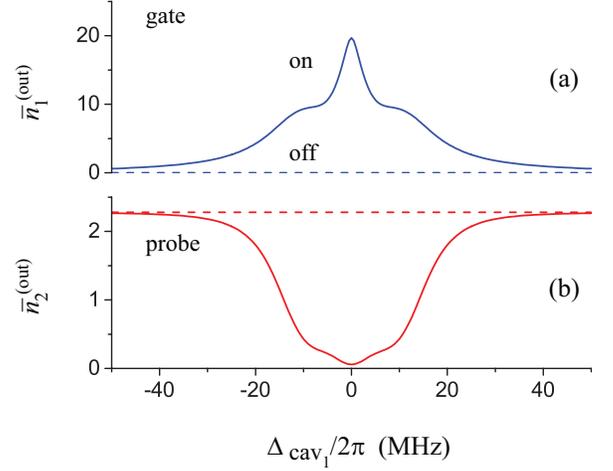}
\end{center}
\caption{(Color online)
Dependencies of the mean output photon numbers $\bar{n}_1^{(\mathrm{out})}$ (a) and $\bar{n}_2^{(\mathrm{out})}$ (b) on the detuning $\Delta_{\mathrm{cav}_1}$ of the gate driving pulse $\mathcal{E}_1$ with respect to the resonant frequency of cavity mode 1 in the cases where cavity mode field 1 is on (solid lines) or off (dashed lines). 
The outputs are integrated from the beginning of the interaction process to the time $t=3$ $\mu$s, which is 1 $\mu$s after the peak time of the probe driving pulse $\mathcal{E}_2$. 
Other parameters are as for Figs.~\ref{fig14} and \ref{fig15}.
}
\label{fig18}
\end{figure}

\begin{figure}[tbh]
\begin{center}
\includegraphics{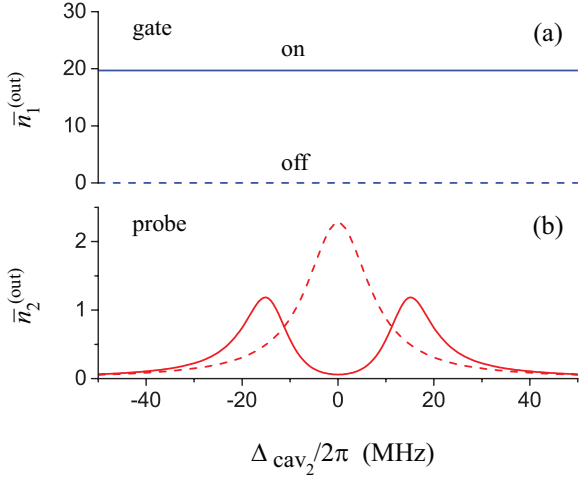}
\end{center}
\caption{(Color online)
Dependencies of the mean output photon numbers $\bar{n}_1^{(\mathrm{out})}$ (a) and $\bar{n}_2^{(\mathrm{out})}$ (b) on the detuning $\Delta_{\mathrm{cav}_2}$ of the probe driving pulse $\mathcal{E}_2$ with respect to the resonant frequency of cavity mode 2 in the cases where cavity mode field 1 is on (solid lines) or off (dashed lines). 
The outputs are integrated from the beginning of the interaction process to the time $t=3$ $\mu$s, which is 1 $\mu$s after the peak time of the probe driving pulse $\mathcal{E}_2$. 
Other parameters are as for Figs.~\ref{fig14} and \ref{fig15}.
}
\label{fig19}
\end{figure} 

We plot in Figs.~\ref{fig18} and \ref{fig19} the dependencies of the mean output photon numbers $\bar{n}_1^{(\mathrm{out})}$ and $\bar{n}_2^{(\mathrm{out})}$
on the detuning $\Delta_{\mathrm{cav}_1}$ of the gate driving pulse $\mathcal{E}_1$ (with respect to the resonant frequency of cavity mode 1) and the detuning $\Delta_{\mathrm{cav}_2}$ of the probe driving pulse $\mathcal{E}_2$ (with respect to the resonant frequency of cavity mode 2)
in the cases where cavity gate mode field 1 is on (solid lines) or off (dashed lines). 
The blue solid curve in Fig.~\ref{fig18}(a) shows that a nearly perfect transparency and a notably narrowed spectrum are obtained in the frequency dependence of the output of cavity mode field 1. These features are due to 
the cavity EIT effect with one atom and have been observed experimentally \cite{Muecke2010}.
Figure~\ref{fig18}(b) shows that the field in probe mode 2 is suppressed by the field in gate mode 1 in a wide region of the detuning $\Delta_{\mathrm{cav}_1}$. The size of this region is determined by the vacuum Rabi frequency of mode 1, that is, by the coupling coefficient $|g_1|/2\pi\simeq 9.4$ MHz. The red solid curve in Fig.~\ref{fig19}(b) shows that there are two peaks which are the signature of the vacuum Rabi splitting of the cavity transmission spectrum of probe mode 2 \cite{Suzuki2011a}.

\begin{figure}[tbh]
\begin{center}
\includegraphics{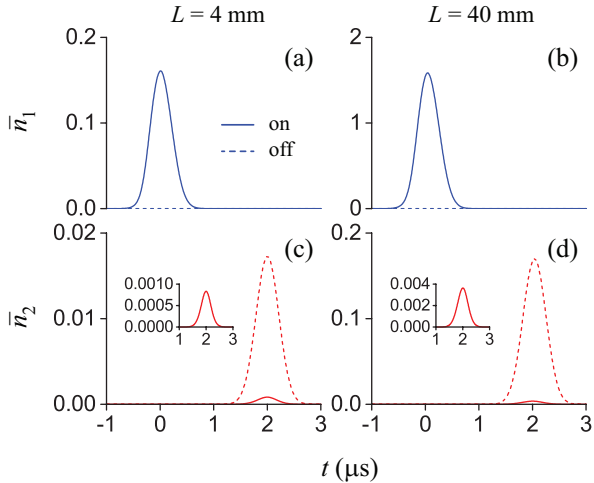}
\end{center}
\caption{(Color online)
Switching of the field in mode 2 by using the field in mode 1 for 
the cavity length $L=4$ mm (left column) and $L=40$ mm (right column).
The solid and dashed curves correspond to the cases where the field in gate mode 1 is on and off, respectively.
Other parameters are as for Figs.~\ref{fig14} and \ref{fig15}.
}
\label{fig20}
\end{figure} 

\begin{figure}[tbh]
\begin{center}
\includegraphics{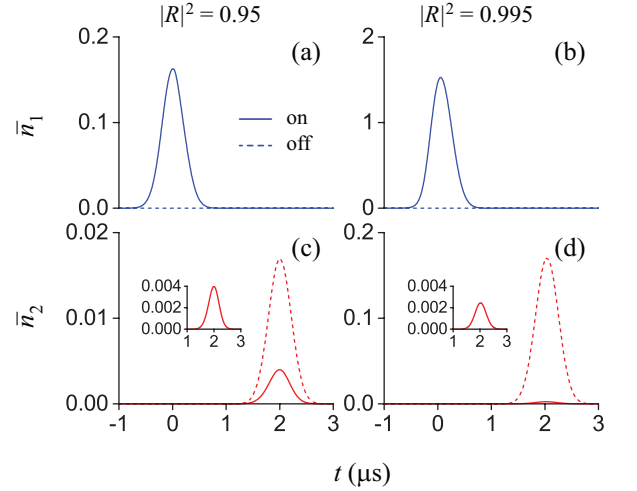}
\end{center}
\caption{(Color online)
Switching of the field in mode 2 by using the field in mode 1 for 
the FBG mirror reflectivity $|R|^2=0.95$ (left column) and $|R|^2=0.995$ (right column).
The solid and dashed curves correspond to the cases where the field in gate mode 1 is on and off, respectively. Other parameters are as for Figs.~\ref{fig14} and \ref{fig15}.
}
\label{fig21}
\end{figure} 

We illustrate in Figs.~\ref{fig20} and \ref{fig21} the switching operations for two different values of the cavity length $L$
and two different values of the FBG mirror reflectivity $|R|^2$, respectively. 
Comparison between the left and right columns of the figures shows that, similar to the results of the previous section, 
the mean cavity-mode photon numbers $\bar{n}_1$ and $\bar{n}_2$ in the right column, 
where $L=40$ mm in the case of Fig.~\ref{fig20} and $|R|^2=0.995$ ($\mathcal{F}\simeq 627$) in the case of Fig.~\ref{fig21},
are larger than those in the left column, where $L=4$ mm in the case of Fig.~\ref{fig20} and $|R|^2=0.95$ ($\mathcal{F}\simeq 61$) in the case of Fig.~\ref{fig21}.
The reason is that, as already stated in the previous section, the values of the damping rate $\kappa$ in the case of the right column is smaller 
than that in the case of the left column. We observe from Figs.~\ref{fig20} and \ref{fig21} that, similar to the results of the previous section, 
the suppression of the mean number $\bar{n}_2$ of photons in the probe mode
increases with increasing cavity length or increasing mirror reflectivity. 
We note that, similar to the results of the previous section, the effect of the cavity length $L$ on the suppression factor is weak, while the effect of the reflectivity $|R|^2$ on the suppression factor is strong. These features are the consequence of the fact that the suppression of the probe field is mainly determined by the cooperativity parameter 
$\eta_2=4|g_2|^2/\gamma_0\kappa$, which depends on $|R|^2$ but does not depend on $L$.

\begin{figure}[tbh]
\begin{center}
\includegraphics{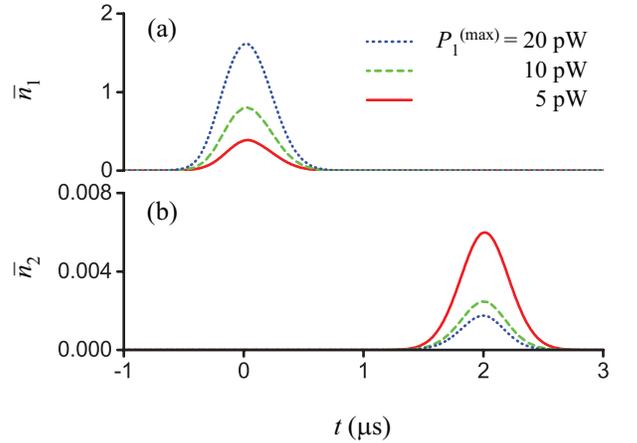}
\end{center}
\caption{(Color online)
Switching of the field in mode 2 by using the field in mode 1 for 
different values of $P_1^{(\mathrm{max})}$. 
The solid and dashed curves correspond to the cases where the field in gate mode 1 is on and off, respectively.
Other parameters are as for Figs.~\ref{fig14} and \ref{fig15}.
}
\label{fig22}
\end{figure} 

We show in Fig.~\ref{fig22} the switching operations for different values of the peak power $P_1^{(\mathrm{max})}$ of the
driving pulse for cavity gate mode 1. We observe from the figure that the suppression of the mean photon number of cavity probe mode 2
increases with increasing power $P_1^{(\mathrm{max})}$. This increase of the suppression factor is a result of the increase in the efficiency of the transfer of the atomic population
from the level $|1\rangle$ to the level $|2\rangle$. Comparison between the curves of Fig.~\ref{fig22}(b) shows that, when $P_1^{(\mathrm{max})}$ is high enough,
the effect of an increase in $P_1^{(\mathrm{max})}$ on $\bar{n}_2$ is not significant. This feature is a consequence of the saturation of the population transfer.

\begin{figure}[tbh]
\begin{center}
\includegraphics{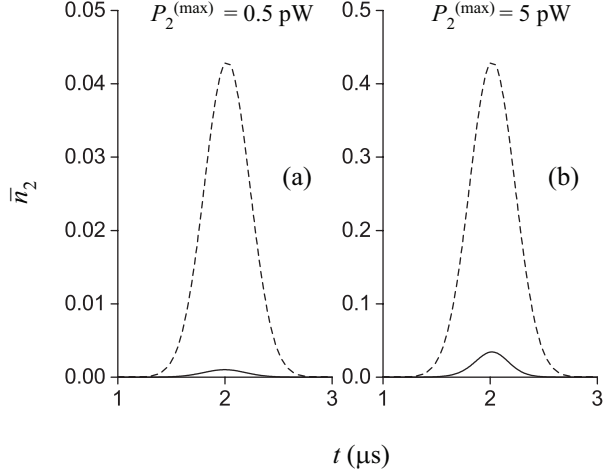}
\end{center}
\caption{
Switching of the field in mode 2 by using the field in mode 1 for different values of $P_2^{(\mathrm{max})}$. 
The solid and dashed curves correspond to the cases where the field in gate mode 1 is on and off, respectively. Other parameters are as for Figs.~\ref{fig14} and \ref{fig15}.
}
\label{fig23}
\end{figure} 

We show in Fig.~\ref{fig23} the time evolution of the mean number $\bar{n}_2$ of photons in cavity probe mode 2 
for different values of the peak power $P_2^{(\mathrm{max})}$ of the driving pulse for this cavity mode. 
We observe from the figure that the suppression of the mean photon number of cavity probe mode 2
decreases with increasing power $P_2^{(\mathrm{max})}$. This decrease of the suppression factor is a result of the power broadening effect.

\begin{figure}[tbh]
\begin{center}
\includegraphics{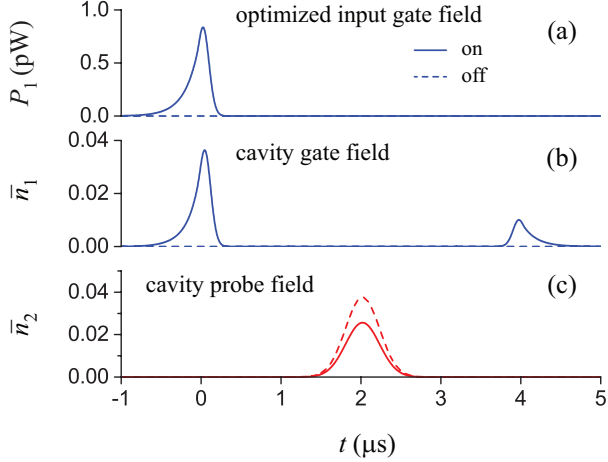}
\end{center}
\caption{
Switching of the field in mode 2 by using the field in mode 1 in the case where the input fields $\mathcal{E}_1$ and $\mathcal{E}_2$ for cavity modes 1 and 2, respectively, are single-photon-energy classical pulses.
The shape of the input pulse for gate mode 1 is optimized by using the procedure of Ref.~\cite{Gorshkov07} and is shown in (a). The solid and dashed curves correspond to the cases where the field in gate mode 1 is on and off, respectively.
Other parameters are as for Figs.~\ref{fig14} and~\ref{fig15}.
}
\label{fig24}
\end{figure}

We show in Fig.~\ref{fig24} the switching operation in the case where the driving pulses $\mathcal{E}_1$ and $\mathcal{E}_2$ for cavity modes 1 and 2, respectively, are single-photon-energy classical pulses, whose energies are equal to the energy of a single photon. Other parameters are as for Figs.~\ref{fig14} and~\ref{fig15}.
We note that a single-photon-level classical pulse, whose pulse energy is equal to or smaller than the energy of a single photon, can be produced by attenuating a coherent laser pulse. The probability of having two photons in such a pulse is small compared to the probability of having one or no photon. When we discard the events with no photon involved, the results can be considered as approximate results for single-photon pulses. 

In order to achieve the maximum efficiency of the storage of light in the atom and consequently the maximum efficiency of the switching, the shape of the input pulse for gate mode 1 is optimized [see Fig.~\ref{fig24}(a)] by using the procedure of Ref.~\cite{Gorshkov07}. This optimization procedure is based on successive time-reversal iterations. When applied to an ensemble of atoms in a traveling-wave ring cavity or in free space, the steps of the procedure are as follows. The atoms are initially prepared in the level $|1\rangle$. Then, for a given writing control field $\mathcal{E}_c^{\mathrm{(write)}}$, a trial input pulse $\mathcal{E}_1^{\mathrm{(in)}}$ is mapped into the spin wave of the lower-level coherence $\rho_{21}$ of the atoms. Both $\mathcal{E}_c^{\mathrm{(write)}}$ and $\mathcal{E}_1^{\mathrm{(in)}}$ are taken to be nonzero over the time interval $(-T_0,0)$. After a storage period $T$, a reading control field $\mathcal{E}_c^{\mathrm{(read)}}(t)=\mathcal{E}_c^{\mathrm{(write)}*}(T-t)$, which is a time-reversed version of the writing control field $\mathcal{E}_c^{\mathrm{(write)}}(t)$, is used to map the lower-level coherence $\rho_{21}$ back into an output pulse $\mathcal{E}_1^{\mathrm{(out)}}$. The input pulse for the next iteration is then generated with a shape corresponding to a time-reversed version of the previous output pulse, and with an amplitude normalized to a fixed energy. These steps are repeated iteratively, using the same writing and reading control fields, until the shape of the output pulse $\mathcal{E}_1^{\mathrm{(out)}}(t)$ coincides with the time-reversed profile $\mathcal{E}_1^{\mathrm{(in)}*}(T-t)$ of the corresponding input pulse $\mathcal{E}_1^{\mathrm{(in)}}(t)$. We extend the above procedure to the case of a single atom in a cavity. We use the control field and the time sequence of Fig.~\ref{fig14}(a). The trial input gate pulse $\mathcal{E}_1^{\mathrm{(in)}}$ has a Gaussian shape with the same full width at half maximum $T_1=0.5$ $\mu$s as that of the gate driving pulse in Fig.~\ref{fig14}(b), but with a peak power $P_1^{(\mathrm{max})}\simeq0.44$ pW, which makes the energy of the pulse equal to that of a single photon. The optimized intensity profile of the input gate pulse $\mathcal{E}_1$ is shown in Fig.~\ref{fig24}(a). The corresponding profiles of the mean photon numbers of the gate and probe modes are shown in Figs.~\ref{fig24}(b) and \ref{fig24}(c), respectively. In Fig.~\ref{fig24}(b), the first and second pulse structures correspond to the writing and reading stages, respectively. Comparison between the solid (gate-on) and dashed (gate-off) curves of Fig.~\ref{fig24}(c) shows that a weak driving pulse with energy of a single photon for cavity mode 1 can suppress the transmission of a driving pulse with energy of a single photon for cavity mode 2. The ratio between the peak values of the mean number $\bar{n}_2$ of photons in probe mode 2 in the gate-off and gate-on cases is equal to about $1.47$. The corresponding value of the switching contrast is about 19\%. Thus, the probe suppression and consequently the switching contrast are not small but not significant.

The efficiency of storage is defined as the ratio between the number of stored excitations, which is given in the case of a single atom by the value of $\rho_{22}$ at the end of the writing stage, and the number of incoming photons. Our additional calculation for $\rho_{22}$ in the case of Fig.~\ref{fig24} shows that the value of the storage efficiency is $f\simeq 0.32$. 
This value is substantially higher than the value of $0.04$ for the storage efficiency in the case of Figs.~\ref{fig14}--\ref{fig17}. However, it is smaller than the limiting optimal value $f_{\mathrm{max}}^{\mathrm{1s}}\simeq 0.81$, which is obtained from the formula $f_{\mathrm{max}}^{\mathrm{1s}}=\eta_1/(1+\eta_1)$ \cite{Gorshkov07}. Here, $\eta_1\simeq 4.3$ is the cooperativity parameter for cavity mode 1. The difference between our obtained value $f\simeq 0.32$ and the limiting optimal value $f_{\mathrm{max}}^{\mathrm{1s}}\simeq 0.81$ arises from the fact that the formula $f_{\mathrm{max}}^{\mathrm{1s}}=\eta_1/(1+\eta_1)$ is valid for an ensemble of atoms  in a traveling-wave ring (one-sided) cavity under the assumption that most of the atoms are in their ground state $|1\rangle$ at all times. Meanwhile, the estimate $f\simeq 0.32$ stands for a single atom with a significant stored excitation magnitude $\rho_{22}$ in a symmetric Fabry-P\'{e}rot (two-sided) cavity. Our additional calculations show that, when we reduce the peak power $P_1^{(\mathrm{max})}$ to a value on the order of or smaller than 1 fW  and perform the optimization procedure, we obtain the optimal storage efficiency $f\simeq 0.4$. This value is in agreement with the estimate 
$f_{\mathrm{max}}^{\mathrm{2s}}=f_{\mathrm{max}}^{\mathrm{1s}}/2=\frac{1}{2}\eta_1/(1+\eta_1)\simeq 0.4$ for the optimal storage efficiency using atoms in a symmetric Fabry-P\'{e}rot cavity with the cooperativity parameter $\eta_1\simeq 4.3$. Note that, when $\eta_1\to\infty$, we have $f_{\mathrm{max}}^{\mathrm{2s}}\to 50\%$.
It is clear that we can improve the optimal storage efficiency by using a one-sided cavity or
an asymmetric Fabry-P\'{e}rot cavity instead of an asymmetric one, and also by increasing the cooperativity parameter $\eta_1$.

\section{Summary}
\label{sec:summary}
We have studied all-optical switches operating on a single four-level atom with the $N$-type transition configuration 
in a two-mode nanofiber cavity with a significant length (on the order of $20$ mm) and a moderate finesse (on the order of 300) under the EIT conditions. In our model, both the gate field and the target field are the quantum nanofiber cavity fields excited by weak classical pulses, and the parameters of the $D_2$ line of atomic cesium are used. We have presented the analytical expressions for the dressed states of the coupled atom-cavity system. We have examined two different schemes for the switching operations. The first scheme is based on the effect of the presence of a photon in the gate mode on the EIT conditions for the probe mode. The second scheme is based on the use of EIT to store a photon of the gate mode in the population of an appropriate atomic level, which leads to the reduction of the transmission of the field in the probe mode. We have investigated the dependencies of the switching contrast on various parameters, such as the cavity length, the mirror reflectivity, and the detunings and powers of the cavity driving field pulses. We have shown that, for a nanofiber cavity with fiber radius of 250 nm, cavity length of 20 mm, and cavity finesse of 313 and a cesium atom at a distance of 200 nm from the fiber surface, it is possible to achieve a switching contrast on the order of about 67\% in the first scheme and of about 95\% in the second scheme. These switching operations require small mean numbers of photons in the nanofiber cavity gate and probe modes. The advantage of the nanofiber-based all-optical switches is that these switches do not require high-finesse cavities. In addition, the nanofiber cavity modes are integrated into the guided modes of the fibers. Consequently, nanofiber-based all-optical switches can find potential applications for quantum information processing and quantum communication networking.

\begin{acknowledgments}
F.L.K. acknowledges support by the Austrian Science Fund (Lise Meitner Project No. M 1501-N27)
and by the European Commission (Marie Curie IIF Grant No. 332255). 
\end{acknowledgments}


\end{document}